\documentclass[conference]{IEEEtran}
\IEEEoverridecommandlockouts
\usepackage{color}
\usepackage{xcolor}
\usepackage{cite}
\usepackage{graphicx}
\usepackage{amsmath,amssymb,amsfonts}
\usepackage{algorithmic}
\usepackage{graphicx}
\usepackage{hyperref}
\usepackage{booktabs}
\usepackage{textcomp}
\usepackage{multirow}
\usepackage{array}
\usepackage{tabularx}

\usepackage{arydshln}

\usepackage[ruled]{algorithm2e}

\usepackage[explicit]{titlesec}

\titlespacing{\section}{0pt}{1.2ex plus .0ex minus .0ex}{.3ex plus .0ex}
\titlespacing{\subsection}{0pt}{1.2ex plus .0ex minus .0ex}{.3ex plus .0ex}

\def\BibTeX{{\rm B\kern-.05em{\sc i\kern-.025em b}\kern-.08em
    T\kern-.1667em\lower.7ex\hbox{E}\kern-.125emX}}
\begin{document}

\title{\color{black}\emph{ACME}: Adaptive Customization of Large Models via  Distributed Systems}

\author{\IEEEauthorblockN{
Ziming Dai,
Chao Qiu, 
Fei Gao,
Yunfeng Zhao,
Xiaofei Wang
}
\IEEEauthorblockA{College of Intelligence and Computing, Tianjin University, Tianjin, China}

\text{Email: \{phoenixdai, chao.qiu, g\_f, yfzhao97, xiaofeiwang\}@tju.edu.cn}
}

\maketitle

\begin{abstract}

Pre-trained Transformer-based large models have revolutionized personal virtual assistants, but their deployment in cloud environments faces challenges related to data privacy and response latency. Deploying large models closer to the data and users has become a key research area to address these issues. However, applying these models directly often entails significant difficulties, such as model mismatching, resource constraints, and energy inefficiency. Automated design of customized models is necessary, but it faces three key challenges, namely, \emph{the high cost of centralized model customization, imbalanced performance from user heterogeneity, and suboptimal performance from data heterogeneity.} In this paper, we propose \textit{ACME}, an adaptive customization approach of Transformer-based large models via distributed systems. To avoid the low cost-efficiency of centralized methods, \textit{ACME} employs a bidirectional single-loop distributed system to progressively achieve fine-grained collaborative model customization. In order to better match user heterogeneity, it begins by customizing the backbone generation and identifying the Pareto Front under model size constraints to ensure optimal resource utilization. Subsequently, it performs header generation and refines the model using data distribution-based personalized architecture aggregation to match data heterogeneity. 
Evaluation on different datasets shows that \textit{ACME} achieves cost-efficient models under model size constraints. Compared to centralized systems, data transmission volume is reduced to 6\%. Additionally, the average accuracy improves by 10\% compared to the baseline, with the trade-off metrics increasing by nearly 30\%.
\end{abstract}


\section{Introduction}

Recently, pre-trained Transformer-based large models, such as Vision Transformer (ViT)\cite{dosovitskiy2020image}, BERT\cite{devlin2019bert}, LLaMA\cite{min2023recent}, and GPT-4\cite{zhai2024large}, have demonstrated exceptional efficiency across diverse tasks. Traditionally deployed in cloud environments, these models face significant challenges with data privacy and response latency, particularly in real-time applications where delays can severely impact user experience and operational efficiency. \textcolor{black}{Therefore, deploying large models on high-performance devices close to users in various vertical domains (e.g., small servers in school laboratories) has become a pivotal research area\cite{zhang2024edgeshard, liu2024mobilellm}. These devices often possess large amounts of domain-specific data, and deploying large models locally can protect data privacy through localized processing and reduced data transmission distances, while also improving system responsiveness.}

However, directly deploying existing lightweight general large models to devices often yields suboptimal performance, as these models typically align poorly with the attributes of heterogeneous devices, including constrained computational resources, diverse data contexts, and stringent energy efficiency requirements \cite{li2021talk, wang2024dependency}. \textcolor{black}{Moreover, manually designing model architectures to match the storage, performance constraints, and data distributions of numerous heterogeneous devices is impractical \cite{liu2023finch}.} To address these problems, there is a critical need to automatically design models that are specifically optimized for the unique operational demands of devices. This approach promises to enhance the efficiency of model deployment in distributed scenarios.

\begin{figure}[t]
	\centering
    \setlength{\abovecaptionskip}{-0.1em} 
   \setlength{\belowcaptionskip}{-0.1em} 
	\includegraphics[width=\linewidth]{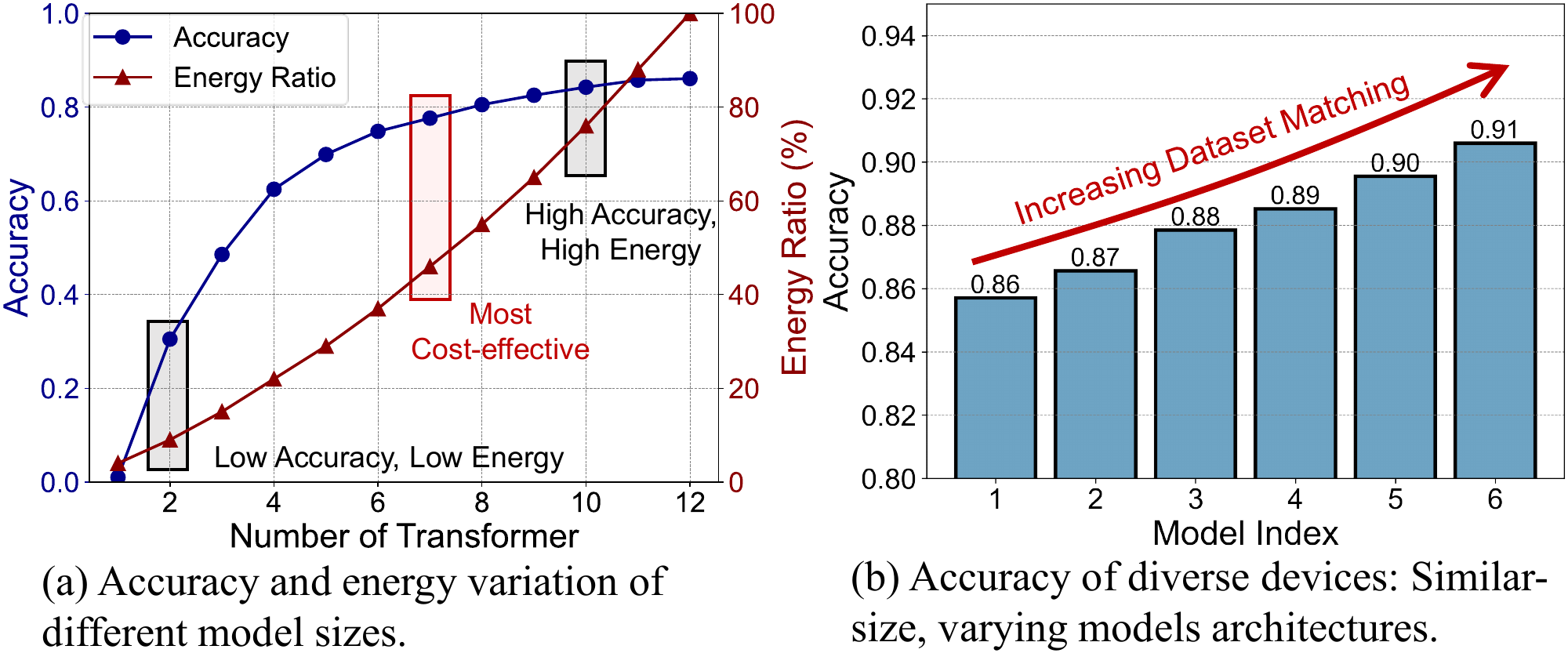}
	\caption{Investigating the relationship among model size, fine-grained architecture, and accuracy: an experimental analysis on CIFAR-100 dataset.}
	\label{fig: motivation}
  \vspace{-1.5em}
\end{figure}

To explore the impact of deploying models on heterogeneous devices, we experiment with the ViT model on the CIFAR-100 \cite{krizhevsky2009learning} dataset, as shown in Fig. \ref{fig: motivation}. Our findings indicate that increasing the model size does not necessarily correlate with performance gains and often leads to higher energy consumption. \textcolor{black}{This result highlights \textbf{the necessity of customizing the most cost-efficient models for devices under specific resource and energy constraints.}} Moreover, comparisons among models of similar model sizes deployed on heterogeneous devices reveal variations in model architecture that result in up to 4.9\% differences in accuracy. This result demonstrates \textbf{the importance of customizing models to fit the unique data distributions in distributed scenarios.}

\begin{figure}[t]
	\centering
	\includegraphics[width=0.9\linewidth]{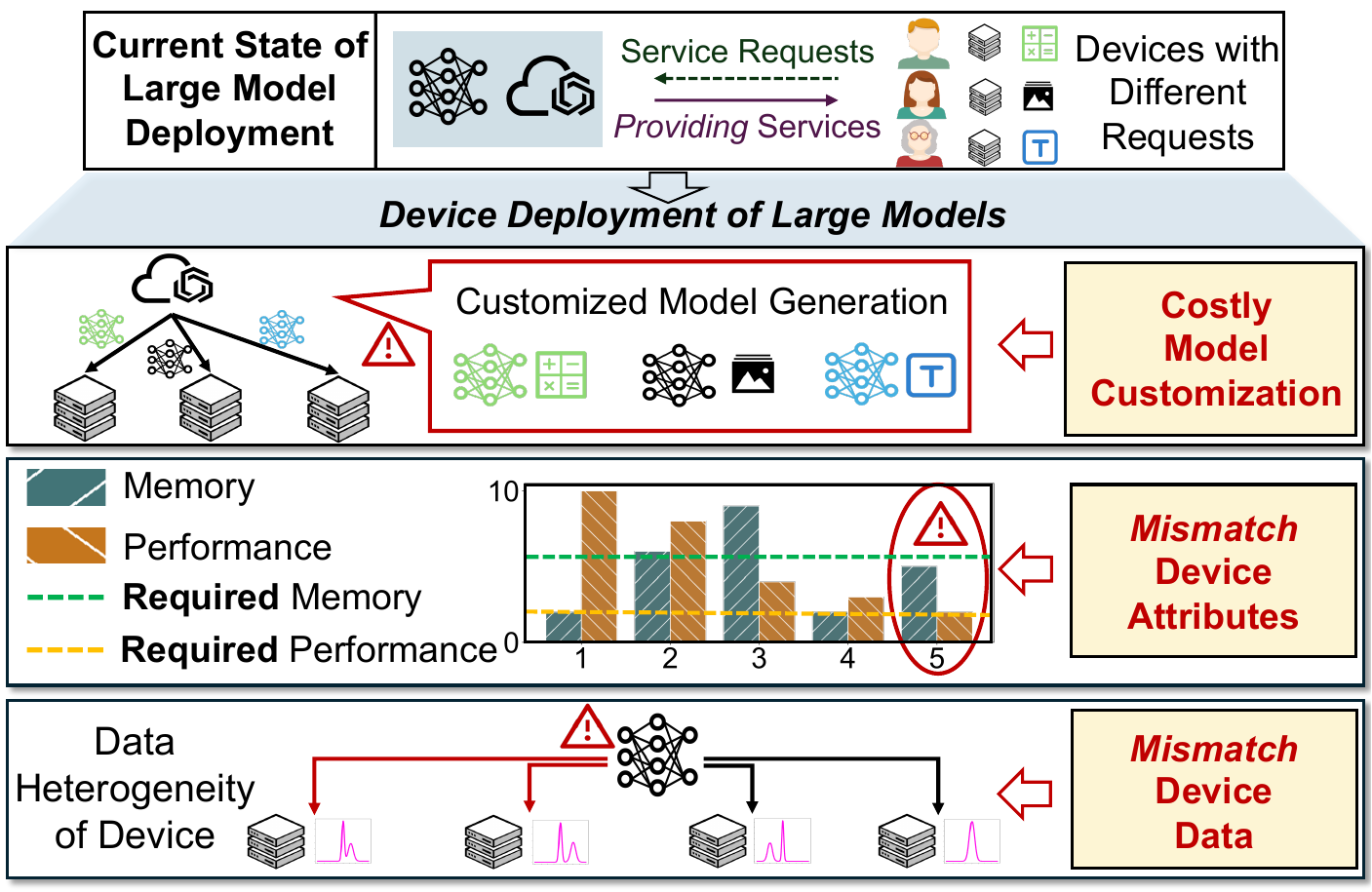}
	\caption{The challenges brought by traditional large model deployment.}
	\label{fig: challenge}
  \vspace{-1em}
\end{figure}

Some previous work has proposed methods to reduce model storage and computational overhead in order to achieve model matching on the device. Techniques used in convolutional neural networks (CNN) (weight pruning \cite{huang2023distributed}, early-exit \cite{ilhan2023scalable}, etc.) are independently applied to Transformer-based models. However, these methods often focus only on the trade-off between model size and performance to obtain a general model. The inherent contradiction between general models and heterogeneous devices in terms of both attributes and data poses significant challenges for ensuring good performance across all devices. \textcolor{black}{Therefore, it is essential to customize models for each device. Nevertheless, it is always difficult due to the three challenges, as shown in Fig. \ref{fig: challenge}:}

{\textit{\textbf{(i) Low cost-efficiency of traditional systems for model customization.}} 
Existing centralized systems (CSs) often struggle to provide fine-grained model customization due to the inherent limitations of the unified customization process in the cloud. This process typically involves significant data transfer between devices and the cloud, as well as substantial operational overhead. The lack of an efficient system makes it cost-inefficient to create customized models that optimally balance performance and resource utilization.
}

\textit{\textbf{(ii) Imbalance performance due to device heterogeneity.}} The heterogeneity in capabilities and energy consumption profiles of devices can lead to imbalanced performance when deploying a model across different devices. Specifically, a model optimized for one device may perform suboptimally or have poor energy efficiency on another device with different characteristics. Moreover, designing a model based on a single performance or attribute fails to account for the diverse constraints imposed by device heterogeneity, potentially hindering the model's overall performance.

\textit{\textbf{(iii) Suboptimal performance due to data heterogeneity.}} The statistical properties of data generated or owned by different devices can vary significantly, leading to diverse optimal model architectures. Models that fail to account for the unique characteristics of each device's data distribution may not effectively capture the underlying patterns, resulting in suboptimal performance.

In this work, we propose an \textbf{A}daptive \textbf{C}ustomization approach of Transformer-based large \textbf{M}od\textbf{E}ls using a 
bidirectional single-loop distributed system, named \textit{ACME}. The main contributions of this paper are summarized as follows:
\begin{itemize}
    {\color{black}\item \textbf{Bidirectional single-loop distributed system for fine-grained model customization:} 
    We introduce a bidirectional single-loop distributed system that enables step-by-step, fine-grained model customization for heterogeneous devices. To enhance the system's high cost-efficiency, we strategically distribute the customization tasks across the cloud server, edge servers, and devices. By leveraging bidirectional information interaction among the three layers, as well as iterative information interaction between the edge servers and devices, we progressively execute model customization from coarse to fine granularity. This approach avoids large-scale data uploads and reduces the overhead, ultimately ensuring optimal model performance and adaptation to device-specific requirements (\S\ref{experi1}).
    }    
    \item \textbf{Attribute-aware customized model matching:
    } \textcolor{black}{We employ a segmentation method based on header and neuron importance to customize the backbone generation, providing various choices to match the device's capacity and storage.} We adopt a grid method to find the Pareto Front under model size constraints, allowing us to quickly obtain the required model architecture for devices after constructing the front. This enhances the matching degree between the model and device attributes, thereby improving the model's trade-off performance (\S\ref{experi3}).
    
    \item \textbf{\color{black}Data-aware personalized model architecture customization:} We perform automatic header generation on different models distributed from the cloud and refine them on local devices. \textcolor{black}{To address the issue of limited data, we introduce a personalized model architecture aggregation technique based on data distribution differences. This technique leverages the knowledge from other devices with similar data distributions to polish the local model architecture, thereby obtaining a model architecture that matches the local data (\S\ref{experi4}).} 
\end{itemize}
\textcolor{black}{Finally, we evaluate \textit{ACME} on different datasets. Results show that \textit{ACME} achieves optimal trade-off models under model size constraints. It requires only 6\% of the data transmission volume compared to CSs, achieving nearly a 10\% improvement in accuracy over the baseline. Furthermore, the trade-off metric, which considers performance, energy consumption, and model size, improves by nearly 30\%.}

\section{System Model}
\subsection{Bidirectional Single-Loop Distributed System Description}

\begin{figure}[t]
	\centering
	\includegraphics[width=\linewidth]{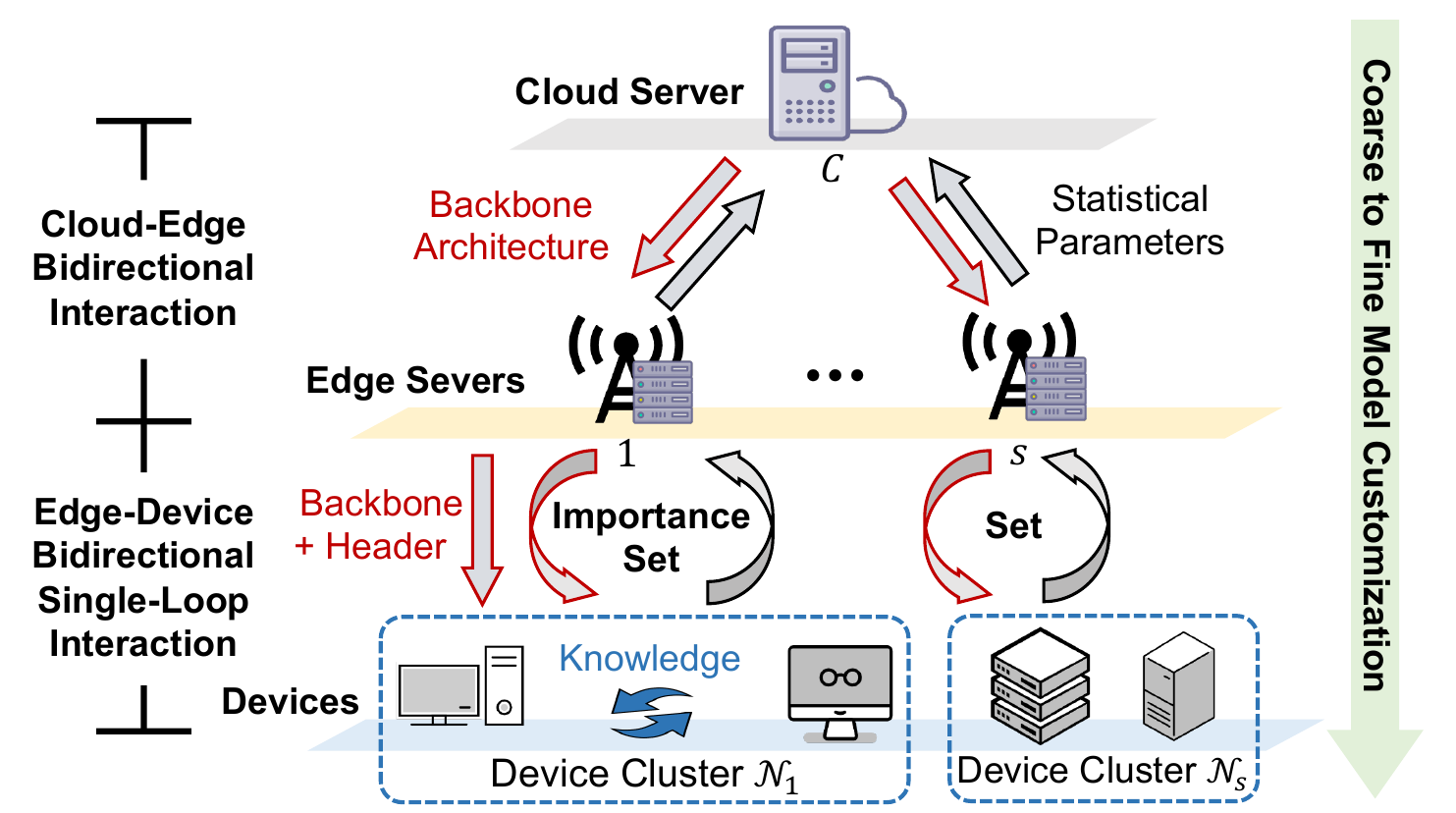}
	\caption{\color{black}The bidirectional single-loop distributed system structure.}
	\label{fig: framework}
  \vspace{-1em}
\end{figure}

\textcolor{black}{We propose a hierarchical bidirectional single-loop distributed system} for generating customized Transformer-based models, with the structure shown in Fig. \ref{fig: framework}. The system architecture is defined as a tuple $(C, \mathcal{S}, \mathcal{N})$, where $C$ represents the cloud server, $\mathcal{S} = \{s_1,  \ldots, s_{|\mathcal{S}|}\}$ denotes the cluster of edge servers, and $\mathcal{N} = \{1, \ldots, N\}$ represents the cluster of devices\cite{zhao2024curcoedge}. 
$S$ and $N$ are the numbers of edge servers and devices, respectively.
The system architecture is characterized by a centralized connection from the cloud server to all edge servers and a partition of devices $\{\mathcal{N}_s \subseteq \mathcal{N} | s=1,\ldots, S\}$, where $\mathcal{N}_s$ is the cluster of devices assigned to edge server $s_s$ \cite{dai2023mg}. 
Each edge server $s_s \in \mathcal{S}$ receives the assigned model from the cloud server and customizes it for its assigned device cluster $\mathcal{N}_s$. The device partitioning is based on similarity in performance and storage capabilities. \textcolor{black}{Given the storage and cost limitations of such devices, we focus on optimizing their model performance, storage, and energy consumption.}

{\color{black}
The system consists of two components of information interaction: the cloud-edge bidirectional interaction and the edge-device bidirectional single-loop interaction, as illustrated in Fig. \ref{fig: framework}:
\begin{itemize}
    \item Cloud-edge bidirectional interaction: This component focuses on backbone customization, achieved through bidirectional information exchange between the cloud server and edge servers. Each edge server $s_s$ first uploads statistical parameters representing the attributes of its device cluster $\mathcal{N}_s$. The cloud server then generates a Pareto Front to match the backbone architecture with the device attributes and subsequently distributes the tailored backbones to the edge servers (Backbone customization).
    \item Edge-device bidirectional single-loop interaction: This component involves header customization, utilizing a bidirectional single-loop information exchange between edge servers and devices. A two-stage \textcolor{black}{customization} process is developed to progressively customize the header architecture. Initially, the edge server generates a header architecture compatible with the distributed backbone using its dataset and distributes the backbone and header to the associated devices. The devices then evaluate different parts of the header architecture based on their local data, identifying importance sets, which are transmitted back to the edge server. The edge server updates the set through personalized architecture aggregation and redistributes them to the devices. This process is repeated iteratively until convergence (Header customization).
\end{itemize}

In summary, bidirectional information interaction spans the entire distributed system, while loop information interaction occurs between edge servers and devices to achieve fine-grained header customization.

}

\subsection{Energy Consumption Model}

To optimize customized model generation, we quantify the energy consumption of Transformer-based models on heterogeneous devices. We approximate the total energy consumption using the backbone's energy consumption. For device $n$ over $k$ epochs, the total energy $E_n(\theta_n)$ is modeled as follows:
\vspace{-0.5em}
\begin{equation}
E_n(\theta_n) = k \cdot \left[P_n(w_n^{\rm B}, d_n^{\rm B}) \cdot T_n(w_n^{\rm B}, d_n^{\rm B})\right],
\label{total_energy_per_epoch}
\vspace{-0.5em}
\end{equation}
where $P_n(\cdot,\cdot)$ and $T_n(\cdot,\cdot)$ represent the power consumption and average latency per epoch, respectively. We define
\vspace{-0.5em}
\begin{equation}
\begin{aligned}
P_n(w_n^{\rm B}, d_n^{\rm B}) &= (G_n + \Delta G_n \cdot w_n^{\rm B}d_n^{\rm B}) + p_n G_n^\beta, \\
T_n(w_n^{\rm B},& d_n^{\rm B}) = (L_n + \Delta L_n \cdot w_n^{\rm B}d_n^{\rm B}), \\
\Delta G_n, &G_n^\beta \propto G_n, \quad \Delta L_n \propto L_n,
\end{aligned}
\label{energy_per_device}
\vspace{-0.5em}
\end{equation}
where $\Delta G_n$ is the power increase per additional Transformer layer, $p_n$ is the number of patches, $G_n^\beta$ is the estimated GPU energy consumption for a batch size $\beta$, and $L_n$ and $\Delta L_n$ represent the initial latency and latency increase per layer, respectively \cite{wang2020energy}. This energy model enables precise optimization of model architectures across heterogeneous devices, balancing performance and energy efficiency.

\subsection{\color{black}Device Attributes and Models}

Assuming devices are located in different geographical locations and handle different tasks, the data between devices should be heterogeneous. We model the heterogeneity of devices and their data as follows: For each device $n \in \mathcal{N}$, we define a tuple $(G_n, C_n, \theta_n)$, where $G_n$ represents GPU capacity. Similar to \cite{tan2019efficientnet}, we use the number of parameters to quantify the model size and device storage constraints. Thus, $C_n$ denotes the maximum storable model parameter count. $\theta_n = (\theta_n^{\rm H}, \theta_n^{\rm B})$ is the customized model, with $\theta_n^{\rm H}$ and $\theta_n^{\rm B}$ representing the header and backbone components, respectively. The model architecture satisfies $|\theta_n^{\rm B}| \gg |\theta_n^{\rm H}|$, where $|\cdot|$ denotes model size. \textcolor{black}{This reflects the typical composition of Transformer-based models, where the backbone stores generalized knowledge for feature extraction, with the multi-layer Transformer architecture at its core. The header maps the features extracted by the backbone to outputs based on the requirements of downstream tasks, typically featuring a simpler network architecture. Consequently, the backbone contains significantly more parameters compared to the header.}

Without loss of generality, we define a reference model $\theta_0=\{\theta_0^{\rm H},\theta_0^{\rm B}\}$ as a baseline. For device $n$, we parameterize its backbone model $\theta_n^{\rm B}$ relative to $\theta_0^{\rm B}$ using a transformation function $\delta(\cdot)$, i.e., $\theta_n^{\rm B} = \delta(\theta_0^{\rm B}, w_n^{\rm B}, d_n^{\rm B})$, where $w_n^{\rm B} \in (0,1]$ represents the width scaling factor and $d_n^{\rm B} \in \mathbb{N}^+$ denotes the number of Transformer layers. It provides a unified framework for analyzing and comparing diverse model architectures, facilitating consistent evaluation across heterogeneous devices.

\subsection{Problem Formulation}

We formulate the problem of generating customized models for heterogeneous devices as a multi-objective optimization problem. The objective is to minimize a composite function that balances model performance, energy consumption, and model size while respecting device-specific storage constraints. Formally, we define the optimization problem as:
\vspace{-0.5em}
\begin{align}
\min_{\left\{ \theta_n, n\in \mathcal{N}\right\} } \
\frac{1}{S}&\sum_{s=1}^S\left\{\frac{1}{|\mathcal{N}_s|}\sum_{n \in \mathcal{N}_s}\left[\mathcal{L}_n(\theta_n, \mathcal{D}_n), E_n(\theta_n), \zeta(\theta_n)\right]\right\},\\
&\text{s.t.}\quad \zeta(\theta_n) \leq C_n,\quad \forall n \in \mathcal{N},\notag
\label{objective_func}
\vspace{-1.5em}
\end{align}
where $\mathcal{L}_n(\theta_n, \mathcal{D}_n)$ is the task-specific loss function for device $n$ with model $\theta_n$ and local data $\mathcal{D}_n$. $\zeta(\theta_n) = d_n^{\rm B}w_n^{\rm B} (H+2\xi_h\xi_f)$ computes the parameter count of $\theta_n$. $H$, $\xi_h$, and $\xi_f$ denote the number of parameters of all heads, hidden layer dimension, and feed-forward layer dimension, respectively.

To address the complexity of our optimization problem, we decompose it into two sub-problems: \textbf{(1) matching the model architecture to device attributes}, and \textbf{(2) matching the model architecture to device-specific data.} Given that $|\theta_n^{\rm B}| \gg |\theta_n^{\rm H}|$, we focus primarily on optimizing the backbone in the first sub-problem. In contrast, the header is adjusted in the second to achieve data-specific adaptation.

This decomposition strategy offers several advantages. The optimized backbone provides effective general representations across diverse datasets without requiring extensive adjustments. Reusing the backbone enhances the model's ability to handle various tasks, improving its generalization capabilities. Header customization captures dataset-specific high-level semantics and category details, improving performance without significantly increasing computational overhead. By separating the optimization of the backbone and header, we achieve a balance between model generalization and task-specific performance while minimizing computational costs. This approach enhances the efficiency and flexibility of our customized model design, enabling effective adaptation to heterogeneous devices and diverse data distributions.

\subsubsection{Phase 1 Problem: Matching the Model Architecture to Device Attributes} 

We formulate the problem of matching model architectures to device attributes as a multi-objective optimization problem. The formula takes into account two attributes of the device, $C_n$ and $G_n$, which respectively affect the model size limit and energy consumption. The objective is to balance model size, energy consumption, and accuracy within device constraints. Formally, we define the optimization problem \textbf{P1} as:
\vspace{-0.7em}
\begin{equation}
 \begin{split}
\text{\textbf{P1:}} \quad   & \min _{\{w_s^{\rm B},d_s^{\rm B}, s=1,\ldots, S\}}    \ \frac{1}{S}\sum_{s=1}^S\left\{\frac{1}{|\mathcal{N}_s|}\sum_{n \in \mathcal{N}_s}f_n(\tilde{\theta}_s)\right\},\\
    & \qquad \text{s.t.} 
    \quad 
    \zeta(\tilde{\theta}_s) < C_n,  \ \forall n \in \mathcal{N},
 \label{objective_func_1}
 \end{split}
 \vspace{-1em}
\end{equation}
where
$f_n(\tilde{\theta}_s) = \left[ \mathcal{L}_n(\tilde{\theta}_s, \tilde{\mathcal{D}}_c), E_n(\tilde{\theta}_s), \zeta(\tilde{\theta}_s) \right]$ and $\tilde{\theta}_s=\left( \theta_0^{\rm H},  \delta(\theta_0^{\rm B},w_s^{\rm B}, d_s^{\rm B}) \right)$.
$\tilde{\mathcal{D}}_c$ represents a generalized public dataset in the cloud server, and $\tilde{\theta}_s$ denotes the intermediate model obtained by $s_s$.

\subsubsection{Phase 2 Problem: Matching the Model Architecture to Device-Specific Data}
After matching the model architecture to device attributes, each edge cluster has a backbone architecture adapted to the attributes of its managed devices, i.e., $\theta_n^{\rm B}=\delta(\theta_0^{\rm B},w_s^{\rm B}, d_s^{\rm B}), n\in\mathcal{N}_s$. We customize the header architecture for each device based on its local data while retaining the cluster-specific backbone. This approach results in a customized model adapted to both device attributes and local data. We formulate the second phase as an optimization problem \textbf{P2}:
\vspace{-0.5em}
\begin{equation}
 \begin{split}
\text{\textbf{P2:}} \quad    \min _{\{\theta_n^{\rm H}, n\in \mathcal{N}\}}  &  \ \frac{1}{S}\sum_{s=1}^S\left\{\frac{1}{|\mathcal{N}_s|}\sum_{n \in \mathcal{N}_s}\mathcal{L}_n(\theta_n, \mathcal{D}_n)\right\},
 \label{objective_func_2}
 \end{split}
 \vspace{-0.5em}
\end{equation}
where $\theta_n=\left\{ \theta_n^{\rm H}, \delta(\theta_0^{\rm B},w_s^{\rm B}, d_s^{\rm B}) \right\},\ n \in \mathcal{N}_s$.
It enables the customization of the header while maintaining the shared optimized backbone architecture within each device cluster managed by the edge server.

\begin{figure}[t]
	\centering
	\includegraphics[width=\linewidth]{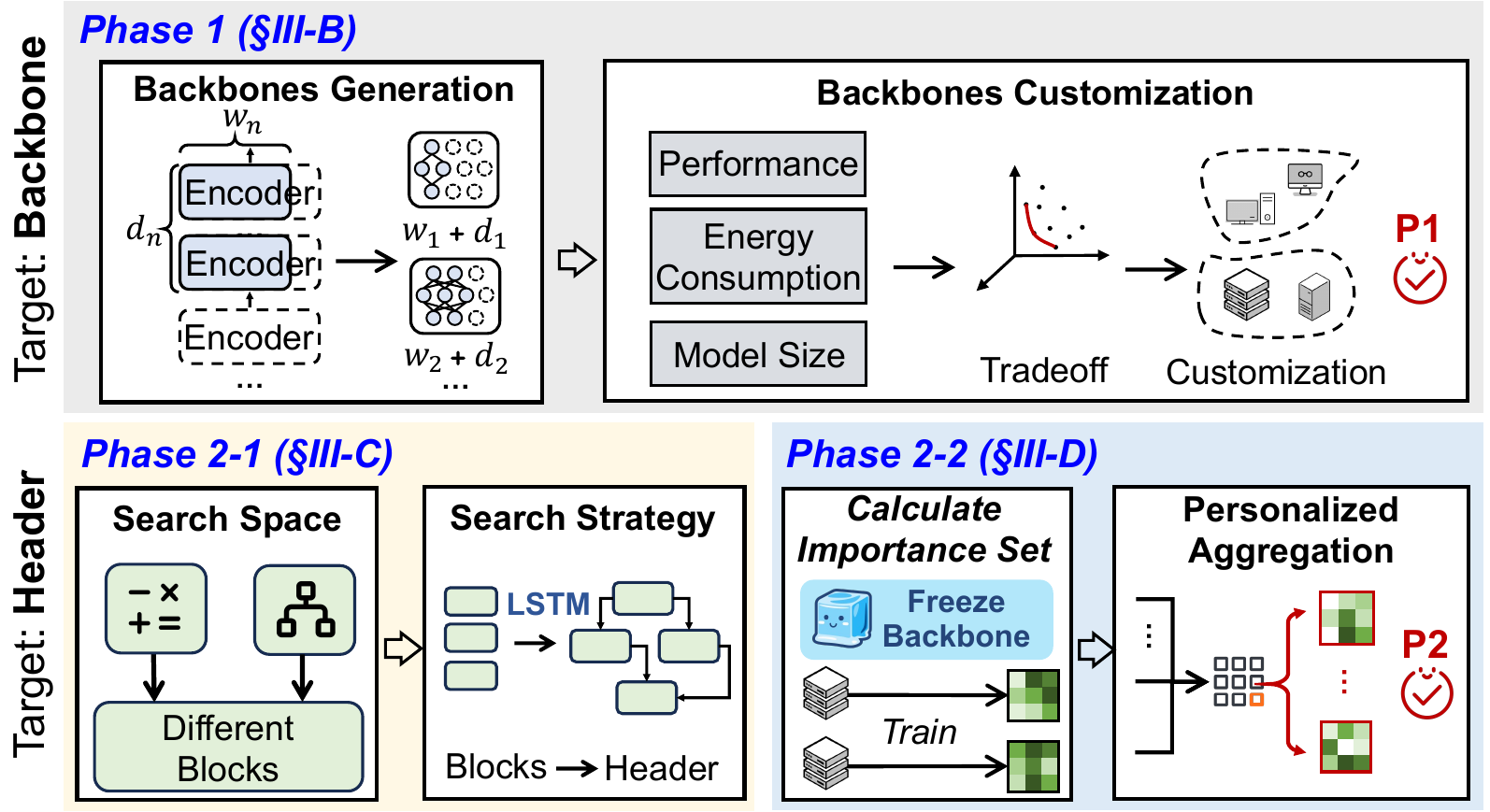}
	\caption{Overall description of the proposed \textit{ACME} process.}
	\label{fig: procedure}
 \vspace{-0.5em}
\end{figure}

\section{The Design of \textit{ACME}}

{\color{black}
\subsection{ACME Overview}
This section provides an overview of the \textit{ACME} process. As illustrated in Fig. \ref{fig: procedure}, our approach adopts a distributed cooperative system to customize personalized model architectures progressively from coarse to fine granularity. The total approach includes two phases and three steps.

\textbf{Backbone customization (\S\ref{p1}).} To address \textbf{P1}, the cloud server performs a two-dimensional segmentation of the model's backbone, generating backbone architectures of various sizes. It then receives the statistical data of device attributes uploaded by each edge server and assigns a backbone architecture to each device cluster $\mathcal{N}_s$, ensuring that the architecture achieves Pareto optimality in terms of performance, model size, and energy consumption within the cluster.

\textbf{Header customization (\S\ref{p2}-\S\ref{p3}).} To tackle \textbf{P2}, we employ two steps. First, the edge server initiates the first-stage customization, leveraging neural architecture search (NAS) on edge server's dataset to generate the header architecture that match the backbone at the block level. Note that the backbone parameters are not frozen during training at this stage. The generated coarse-grained header architecture is then distributed to the connected devices (\S\ref{p2}).

Each device performs the second-stage customization by freezing the backbone architecture and its parameters, training the header using local private dataset, and generating an importance set for the architecture, which is uploaded back to the edge server. The edge server aggregates all importance sets from the device cluster using the Wasserstein distance for personalized architecture aggregation and distributes the customized importance set to each device. This step enables the model to progressively refine from coarse-grained to fine-grained customization, effectively aligning with different devices while leveraging collaboration among devices to optimize local model architectures (\S\ref{p3}).

}

\subsection{Phase 1: Generate and Customize Backbone to Match Devices Attributes}\label{p1}

\subsubsection{Backbone Generation}

To accommodate devices with varying attributes, we propose a systematic approach for generating multiple smaller backbone models from a complete backbone through selective component removal. This process targets redundant elements in both width (embedding dimensions, attention heads, Multi-Layer Perceptron (MLP) dimensions) and depth (Transformer layers). Given the interdependence of two dimensions in our strategy, we adopt a two-step approach. First, we derive an architecture $\acute{\theta}^{\rm B}$ with variable width, which serves as the basis for the final backbone architecture $\theta^{\rm B}$ \cite{hou2020dynabert}.

We quantify the importance of heads in the Multi-Head Self-Attention (MSA) module and neurons in the MLP using a small dataset $\mathcal{D}_C$. The importance metric $I_{h}$ for a head $h$ with output $O_h$ is defined as:
\vspace{-0.5em}
\begin{equation}
I_{h} = |\mathcal{F}(O_h,\mathcal{D}_C) - \mathcal{F}(O_{h=0},\mathcal{D}_C)|,
\label{loss_change}
\end{equation}
where $\mathcal{F}(O_h,\mathcal{D}_C)$ denotes the training loss function of the model with head $h$ present on dataset $\mathcal{D}_C$, and $\mathcal{F}(O_{h=0},\mathcal{D}_C)$ represents the loss function after removing head $h$. $O_{h=0}$ is the output of the model without head $h$. We approximate $\mathcal{F}(O_{h=0},\mathcal{D}_C)$ using a first-order Taylor expansion as follows:
\begin{equation}
\mathcal{F}(O_{h=0},\mathcal{D}_C) = \mathcal{F}(O_h,\mathcal{D}_C) - \frac{\partial \mathcal{F}}{\partial O_h}(O_h-O_{h=0})+R_{h=0},
\label{head_loss_function}
\end{equation}
where $R_{h=0}$ is the remainder term, which can be ignored when $O_h$ is close to $O_{h=0}$ because it is much smaller than the previous terms. When removing head $h$, $O_{h=0}=0$. Therefore, combining Eq. (\ref{loss_change}) and Eq. (\ref{head_loss_function}), we calculate $I_{h}$ as follows:
\begin{equation}
 \begin{split}
    I_{h} = \left|\frac{\partial \mathcal{F}}{\partial O_h}O_h-R_{h=0}\right|\approx \left|\frac{\partial \mathcal{F}}{\partial O_h}O_h\right|.
 \label{head_importance}
 \end{split}
\end{equation}

The importance of neurons can be obtained similarly. Based on these importance metrics, we rank each Transformer's heads and neurons. According to the requirements, we discard those at the bottom of the list to obtain $\acute{\theta}^{\rm B}$, which is the backbone model with dynamically adjustable width $\mathcal{W}^{\rm B}$.

To generate the final backbone architecture $\theta^{\rm B}$ with dynamic width $\mathcal{W}^{\rm B}$ and dynamic depth $\mathcal{D}^{\rm B}$, we employ knowledge distillation using $\acute{\theta}^{\rm B}$ as the teacher model. The distillation objective function is formulated as follows:
\vspace{-0.5em}
\begin{equation}
\mathcal{L}(\acute{\theta}^{\rm B}, {\theta}^{\rm B})= \lambda_1 l(\acute{\textbf{y}}, \textbf{y}) + \lambda_2 l(\acute{\textbf{E}}, \textbf{E}) + l(\acute{\textbf{H}}, \textbf{H}),
\label{distillation_obj}
\end{equation}
where $l(\cdot, \cdot)$ denotes the mean squared error function. The terms $\acute{\textbf{y}}$, $\acute{\textbf{E}}$, and $\acute{\textbf{H}}$ represent the logits, embeddings, and hidden states of the teacher model $\acute{\theta}^{\rm B}$, respectively. Similarly, $\textbf{y}$, $\textbf{E}$, and $\textbf{H}$ correspond to those of the student model $\theta^{\rm B}$. The coefficients $\lambda_1$ and $\lambda_2$ balance the contributions of different components in the distillation process.

\subsubsection{Backbone Customization}

Similar to the constrained decomposition with grids proposed by Cai \textit{et al.}\cite{cai2017constrained}, we decompose each objective function in \textbf{P1} into multiple sub-problems and use a grid approach to find the Pareto Front. We use an approximate method to solve \textbf {P1}, specifically using the maximum energy consumption within each device cluster as the representative metric:
\vspace{-0.6em}
\begin{equation}
 \begin{split}
\min_{\{w_s^{\rm B},d_s^{\rm B}, s=1,\ldots, S\}}&    \ \frac{1}{S}\sum_{s=1}^Sf_s(\tilde{\theta}_s),\\
    \text{s.t.} \quad
    \zeta(\tilde{\theta}_s) < & \min_{n\in \mathcal{N}_s} C_n,
 \label{objective_func_1_1}
 \end{split}
\end{equation}
where $f_s(\tilde{\theta}_s) = \left[ \mathcal{L}_s(\tilde{\theta}_s, \tilde{\mathcal{D}}_c), E_s(\tilde{\theta}_s), \zeta(\tilde{\theta}_s) \right],$
$\tilde{\theta}_s=( \theta_0^{\rm H}, \tilde{\theta}^{\rm B}_s =
$ $
\delta(\theta_0^{\rm B},w_s^{\rm B}, d_s^{\rm B}) ),$ 
and
  $E_s(\tilde{\theta}_s)  = \max_{n\in \mathcal{N}_s} E_n(\tilde{\theta}_s).$
$\mathcal{L}_s(\tilde{\theta}_s, \tilde{\mathcal{D}}_c)$ is the loss function of the model  $\tilde{\theta}_s$ under dataset $\tilde{\mathcal{D}}_c$. For ease of representation, we let $f^l_s(\tilde{\theta}_s), l\in\{1,2,3\}$ denote the $l$th element of $f_s(\tilde{\theta}_s)$.

\begin{algorithm}[t]
\caption{Backbone customization algorithm}
\begin{algorithmic}[1]
\STATE \textbf{Input:} $\mathcal{N}, \theta_0=\{\theta_0^{\rm B}, \theta_0^{\rm H}\}$;
\STATE Compute importance of headers and neurons on $\mathcal{D}_C$;
\STATE Remove $\theta_0^{\rm B}$'s headers and neurons to get $\acute{\theta}^{\rm B}$;
\STATE Obtain $\theta^{\rm B}$ with $\mathcal{D}^{\rm B}$ and $\mathcal{W}^{\rm B}$ by Eq. (\ref{distillation_obj});
\FOR{\text{each device cluster $\mathcal{N}_s \subseteq \mathcal{N}$}}
    \STATE $K \leftarrow |f_s^l(\tilde{\theta}_s^*)-f_s^l(\tilde{\theta}_s^{-})| / \gamma_p$;
    \FOR{\text{each $w^{\rm B} \in \mathcal{W}^{\rm B}$ and $d^{\rm B} \in \mathcal{D}^{\rm B}$}}
        \STATE Initialize $\tilde{\theta}_s = \{\delta(\theta_0^{\rm B},w^{\rm B}, d^{\rm B}), \theta_0^{\rm H}\}$;
        \STATE Compute $ f^l_s(\tilde{\theta}_s), l \in \{1,2,3\} $;
        \FOR{\text{each sub-problem $l$}}
            \STATE $\Psi^l(\tilde{\theta}_s) \leftarrow \left\lceil\left(f_s^l(\tilde{\theta}_s) - f_s^l(\tilde{\theta}_s^*)+\sigma\right) / r^l\right\rceil$;
        \ENDFOR
    \ENDFOR
    \FOR{\text{every $l \in \{1,2,3\}$ and $k \in \{1,\ldots,K\}$}}
        \STATE Construct $ \Phi^{l,k}_s $ such that $\Psi^l(\tilde{\theta}_s) = {g_s^{l,k}}^*$;
    \ENDFOR
    \STATE Obtain the Pareto Front Grid based on $ \Phi^{l,k}_s $;
    \STATE Find the final model $ \tilde{\theta}_s$ by Eq. (\ref{select_Backbone}).
\ENDFOR
\end{algorithmic}
\label{algorithm1}
\end{algorithm}

We introduce a performance window $\gamma_p$ to quantify the acceptable trade-off between performance and other objectives. We partition the objective space into $K = |f_s^l(\tilde{\theta}_s^*)-f_s^l(\tilde{\theta}_s^{-})| / \gamma_p$ intervals between the ideal optimal point $\tilde{\theta}_s^*$ and the worst-case point $\tilde{\theta}_s^{-}$, where $l=1$ corresponds to the performance metric. This partitioning is applied uniformly across all objectives.
For each backbone architecture, we compute grid coordinate $\Psi^l(\tilde{\theta}_s)$ for $l$th objective function as:
 \vspace{-0.5em}
\begin{equation}
 \begin{split}
    \Psi^l(\tilde{\theta}_s)& = \lceil(f_s^l(\tilde{\theta}_s) - f_s^l(\tilde{\theta}_s^*)+\sigma) / r^l\rceil, \\
    r^l =& [f_s^l(\tilde{\theta}_s^{-})-f_s^l(\tilde{\theta}_s^*)+2\sigma]/K,
 \label{grid_func}
 \end{split}
  \vspace{-0.5em}
\end{equation}
where $\sigma > 0$ is a small constant to prevent division by zero.
We define the solution set $S_s^l(k)$ for the  $l$th objective in the $k$th interval as:
\vspace{-0.4em}
\begin{equation}
 \begin{split}
    S_s^l(k)=\{\tilde{\theta}_s|\Psi^1(\tilde{\theta}_s)=1, \ldots,\Psi^{l-1}(\tilde{\theta}_s)=l-1,\\
    \Psi^{l+1}(\tilde{\theta}_s)=l+1,\ldots,\Psi^{K}(\tilde{\theta}_s)=K
    \}.
 \label{sub-problem_func}
 \end{split}
\end{equation}

For each objective function $f_s^l(\tilde{\theta}_s)$ of edge server $s_s$, we define ${g_s^{l,k}}^*$ as the optimal value of the grid coordinates $\Psi^l(\tilde{\theta}_s)$ among all solutions in the $k$th interval $S_s^l(k)$. We then construct $\Phi^{l,k}_s$ as the set of solutions in $S_s^l(k)$ satisfying $\Psi^l(\tilde{\theta}_s) = {g_s^{l,k}}^*$. The union of all $\Phi^{l,k}_s$ forms the Pareto Front Grid (PFG) \cite{Xu2023APF}, which approximates the Pareto-optimal set in the discretized objective space. 

Given the stringent storage constraints above, we propose an efficient approach to Pareto Front construction and model selection. We define $\tilde{\theta}_s^{-}$ such that $\zeta'(\tilde{\theta}_s^{-}) = \min_{n\in \mathcal{N}_s} C_n$, effectively establishing an upper bound on model size. We then construct a truncated PFG by discarding all models exceeding this limit. Within this constrained PFG, we first identify the model with the highest performance. Using the grid location of this high-performing model as our search space, we select the final model $\tilde{\theta}_s$ that minimizes the Euclidean distance to the ideal optimal point $\tilde{\theta}_s^{*}$, i.e., 
\vspace{-0.5em}
\begin{equation}
 \begin{split}
    ({w}_s^{\rm B},{d}_s^{\rm B}) = \arg \min \limits_{(w_s^{\rm B}, d_s^{\rm B}) \in \Phi_s^{\rm h}} \sum_{l=1}^3\|\Psi^l(\tilde{\theta}_s) - \Psi^l(\tilde{\theta}_s^*)\|_2, 
 \label{select_Backbone}
 \end{split}
\end{equation}
where $\tilde{\theta}_s=( \theta_0^{\rm H}, \tilde{\theta}^{\rm B}_s=\delta(\theta_0^{\rm B},{w}_s^{\rm B},{d}_s^{\rm B}) )$ and $\Phi_s^{\rm h}$ represents the set of models within the grid location of the highest-performing model that satisfies the storage constraint. The detail for backbone customization is shown in Algorithm \ref{algorithm1}.

\subsection{Phase 2-1: First-Stage \textcolor{black}{Customization} for Generating Coarse Header of Matching Data}\label{p2}
\begin{figure}[t]
	\centering
	\includegraphics[width=\linewidth]{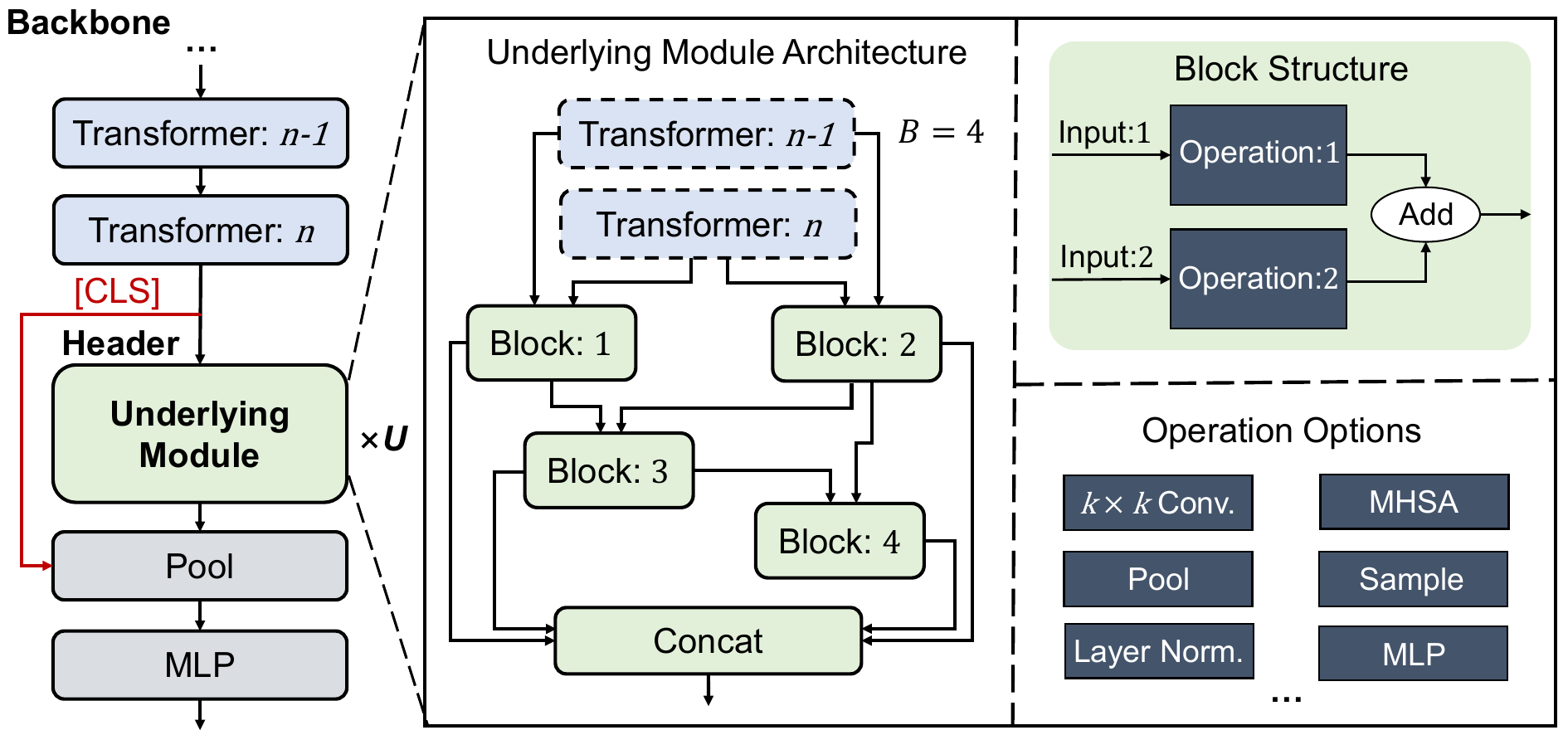}
	\caption{The search space and the architecture that might be searched.}
	\label{fig: searchspace}
  \vspace{-1em}
\end{figure}

\subsubsection{Generating Search Space Based on Blocks}
{\color{black}Due to the indeterminate nature of the backbone architecture deployed by the cloud server, manually designing headers is time-consuming and cannot guarantee optimal performance. As a technology within the auto machine learning domain, NAS defines a search space and strategy to automatically find the best model architecture based on the data. This approach can effectively design a header architecture that is well-suited for each specific backbone. Furthermore, the relatively simple architecture of the headers reduces the complexity of the search process for NAS.}

Inspired by the progressive NAS approach \cite{liu2018progressive}, we formulate the header architecture for each edge server $s_s$ as a composition of predefined operational blocks. The underlying module architecture of the header is represented as a Directed Acyclic Graph (DAG) comprising $B$ blocks, i.e., $\mathcal{B}=\{B_1,\ldots, B_B\}$, which is repeated $N$ times, followed by a pooling layer and an MLP. Each block $B_b$ in the DAG is defined as a 5-tuple $(\hat{I}_{b,1}, \hat{I}_{b,2}, \hat{O}_{b,1}, \hat{O}_{b,2}, \hat{C})$, where $\hat{I}_{b,1}, \hat{I}_{b,2} \in \hat{\mathcal{I}}_b$ denote the input tensors. We define the input set $\hat{\mathcal{I}}_b$ for block $B_b$ as the union of all preceding blocks, the backbone output, and the penultimate Transformer layer output from the backbone. $\hat{O}_{b,1}, \hat{O}_{b,2}\in \hat{\mathcal{O}}$ represent the operations applied to the inputs. The operation set $\hat{\mathcal{O}}$ comprises a diverse range of functions commonly used in model headers, including $z \times z$ convolution layers, pooling layers, and downsampling operations \cite{xu2023lgvit, bakhtiarnia2022single}. $\hat{C}$ specifies the combination method for operation outputs.

Following the findings of \cite{zoph2018learning}, we constrain the combination function $\hat{C}$ to element-wise addition, effectively reducing the search space while maintaining expressive power. To accommodate dimension mismatches, we insert $1 \times 1$ convolution layers as necessary. To leverage the global feature representation captured by the backbone, we incorporate the [CLS] token into the final pooling output of the header. This integration promotes synergy between the global information from the backbone and the local features extracted by the header. Fig. \ref{fig: searchspace} shows the header architecture composed of $U$ underlying modules with $B=4$, demonstrating the potential complexity and flexibility of our proposed search space.

To quantify the search space and demonstrate its capacity to accommodate diverse header requirements for various backbone architectures,  we formulate the following analysis. Let $\hat{\mathbb{B}}_b$ denote the possible architecture space for the $b$th block. The cardinality of $\hat{\mathbb{B}}_b$ is given by: $|\hat{\mathbb{B}}_b| = |\hat{\mathcal{I}}_b|^2 \times |\hat{\mathcal{O}}|^2$, where $|\hat{\mathcal{I}}_b| = (b+1)$. 
Since $U$ does not affect the search space, the total number of possible architectures for a header with $B$ blocks can be expressed as follows, based on \cite{liu2018progressive}:
\vspace{-0.5em}
\begin{equation}
|\hat{\mathbb{B}}_{1:B}| = \prod_{b=1}^B \left( (b+1)^2 \times |\hat{\mathcal{O}}|^2 \right).
\label{blocks_func}
\vspace{-0.5em}
\end{equation}

\subsubsection{Block Selection and Coarse Header Generation}

Edge server $s_s$ employs a controller to determine the block selection and connection strategy for its device cluster, ultimately defining the coarse-grained model $\theta_s$. To accommodate variable-sized inputs, we implement a Long Short Term Memory (LSTM)-based controller, capable of processing sequences of length $4B$. The sequence element $B_b$ represents one of the four components of a block: $\hat{I}_{b,1}, \hat{I}_{b,2}, \hat{O}_{b,1}$ and $\hat{O}_{b,2}$. The input to LSTM is a one-hot encoded vector of size $|\hat{\mathcal{I}}_b|$ or $|\hat{\mathcal{O}}|$, followed by an embedding lookup. The final hidden state of the LSTM is transformed through a fully connected layer and a sigmoid function to estimate the validation's accuracy.

To reduce the search complexity, we adopt a parameter-sharing strategy for child models \cite{10366825}. The search process involves two sets of learnable parameters: the LSTM controller parameters $\boldsymbol{\theta}^{\rm LSTM}_s$ for edge server $s_s$, and the shared parameters $\omega_s$ of the child models. 
The search alternates between two steps:
1) Optimizing $\omega_s$ using the shared dataset;
2) Updating $\boldsymbol{\theta}^{\rm LSTM}_s$ based on the performance of sampled architectures. This alternating optimization continues throughout the search process. In our implementation, the shared parameters $\omega_s$ correspond to the weights of operations within each block, as our approach uses the same layer stacking. This parameter-sharing mechanism significantly reduces the computational overhead of architecture search by allowing multiple child models to reuse the same set of parameters.

During the optimization of shared parameters $\omega_s$ in edge server $s_s$, we fix the controller's policy $\pi_s(\theta^{\rm H}_s,\boldsymbol{\theta}^{\rm LSTM}_s)$, where $\theta^{\rm H}_s$ represents the sampled child model. We minimize the expected loss $\mathbb{E}_{\theta^{\rm H}_s\sim \pi_s(\theta^{\rm H}_s,\omega_s)}[\mathcal{L}^l(\theta^{\rm H}_s,\omega_s)]$ using stochastic gradient descent. Monte Carlo sampling estimates gradient, i.e.,
  \vspace{-0.5em}
\begin{equation}
\nabla_{\omega_s} \mathbb{E}_{\theta^{\rm H}_s\sim \pi_s(\theta^{\rm H}_s,\omega_s)}[\mathcal{L}^l(\theta^{\rm H}_s,\omega_s)] \approx \frac{1}{M}\sum_{i=1}^{M}\nabla_{\omega_s}\mathcal{L}^l(\theta^{\rm H}_{s,i},\omega_s),
\label{monte_func}
\end{equation}
where $\theta^{\rm H}_{s,i}$ is an independent sample from $\pi_s(\theta^{\rm H}_s,\omega_s)$.
For optimizing the controller parameters $\boldsymbol{\theta}^{\rm LSTM}_s$ of edge server $s_s$, we fix $\omega_s$ and maximize the expected reward $\mathbb{E}_{\theta^{\rm H}_s\sim \pi_s(\theta^{\rm H}_s,\omega_s)}[R^l(\theta^{\rm H}_s,\omega_s)]$ \cite{pham2018efficient}. We employ the REINFORCE algorithm \cite{williams1992simple} to compute policy gradients, incorporating a moving average baseline to reduce variance.

\subsection{Phase 2-2: Second-Stage \textcolor{black}{Customization} with Personalized Architecture Aggregation for Generating Fine Header Suitable for Data}\label{p3}
\begin{figure}[t]
	\centering
	\includegraphics[width=\linewidth]{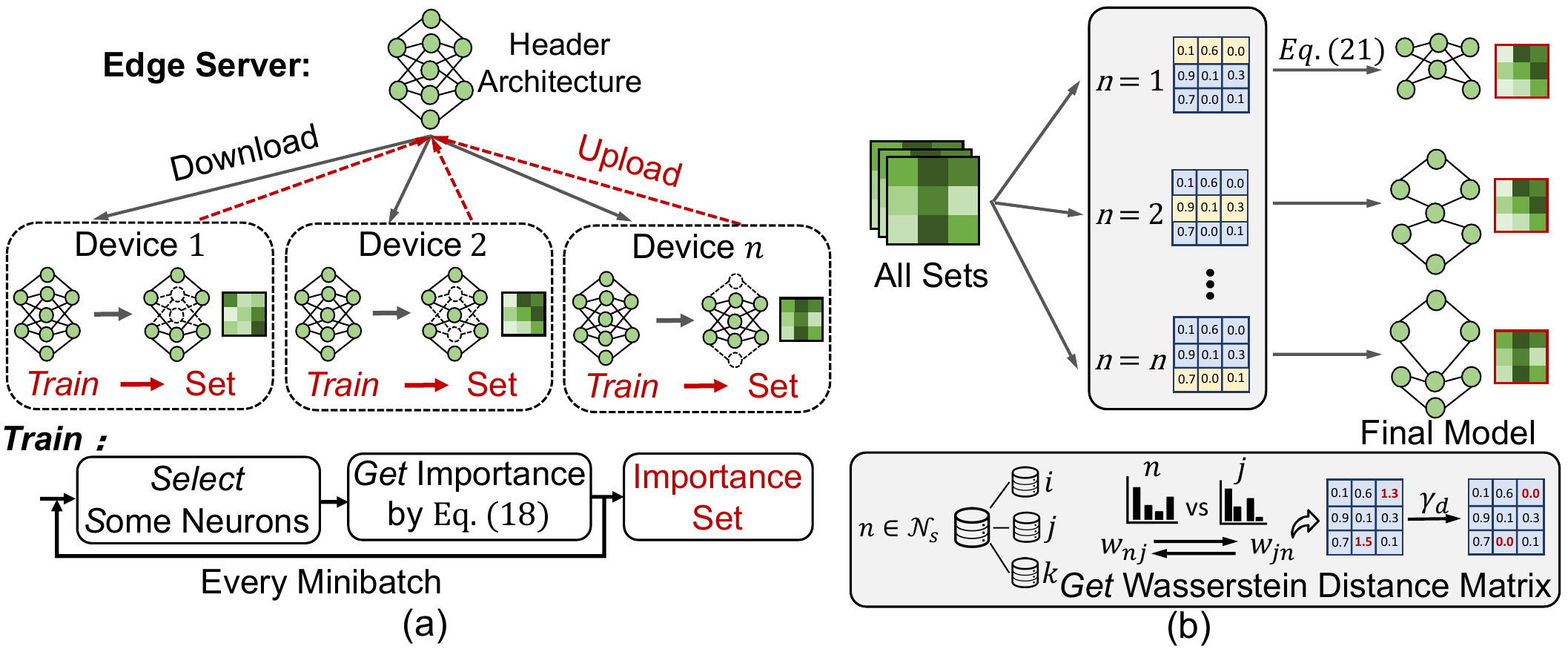}
	\caption{Importance set generation and personalized architecture aggregation.
 }
	\label{fig: prnue}
  \vspace{-1em}
\end{figure}

The header architecture $\theta^{\rm H}_s$, generated by the edge server $s_s$, is coarse-grained.
This architecture may not be optimal for the local data distributions of individual devices. To address this limitation, we propose another two-step process:
1) Device-specific header customization: Each device generates a header optimized for its local data distribution;
2) Collaborative knowledge integration: Devices collaborate with edge servers to leverage collective information to refine their header architectures, enhancing model performance.

\subsubsection{Architecture Quantification Based on Importance}

The studies \cite{hou2020dynabert, molchanov2019importance} highlight that the importance of parameters varies with the task and data, and eliminating unnecessary parameters can improve model performance. In this step, considering the limitations of devices' capabilities, we choose a minimal overhead strategy to measure the importance based on the gradients accessible during training. As shown in Fig.~\ref{fig: prnue} (a), we use the above strategy to quantify the importance of parameters of the coarse-grained header with device $n$'s local data, discarding redundant parts based on the importance set $\mathcal{Q}_n$ and adapting the header to local data.

Device $n \in \mathcal{N}_s$ receives $\hat{\theta}_n=\theta_s$ from the edge server $s_s$, where parameters of the header $\Upsilon^{\rm H}_n = \{\upsilon_{n,1}^{\rm H}, \upsilon_{n,2}^{\rm H},\ldots ,\upsilon_{n,R}^{\rm H}\}$. Thus, $\hat{\theta}_n$ can be written as $\{\hat{\theta}_n^{\rm B}, \Upsilon^{\rm H}_n\}$. Assuming parameters in $\Upsilon^{\rm H}_n$ are independent, we can determine which parameters to remove by considering their importance individually. Starting from the complete $\Upsilon^{\rm H}_n$, we reduce the parameters in $\Upsilon^{\rm H}_n$ one by one, quantifying each parameter by the error introduced by its removal. Based on the independent and identically distributed (IID) assumption, the induced error can be measured by the squared difference in prediction error with and without the parameter $\upsilon^{\rm H}_{n,r}$ \cite{molchanov2019importance}:
\vspace{-0.5em}
\begin{equation}
\textbf{Q}_{n,r}=\left(\mathcal{L}_n(\Upsilon^{\rm H}_n,\mathcal{D}_n)-\mathcal{L}_n(\Upsilon^{\rm H}_n|\upsilon^{\rm H}_{n,r}=0,\mathcal{D}_n)\right)^2.
 \label{importance_param}
 \vspace{-0.5em}
\end{equation}

To avoid repeatedly evaluating $|\Upsilon^{\rm H}_n|$ times when calculating the importance of the $r$th parameter $\textbf{Q}_{n,r}$, we approximate $\textbf{Q}_{n,r}$ near $\Upsilon^{\rm H}_n$ using a first-order Taylor expansion:
\vspace{-0.5em}
\begin{equation}
    \textbf{Q}^{(1)}_{n,r}(\Upsilon^{\rm H}_n)=\left(g_{n,r}\upsilon^{\rm H}_{n,r}\right)^2, \quad
    g_{n,r}=\frac{\partial \mathcal{L}_n}{\partial \upsilon^{\rm H}_{n,r}}.
    \vspace{-0.5em}
 \label{importance_param_taylor}
\end{equation}
Therefore, we get the importance set $\mathcal{Q}_n$:
 \vspace{-0.5em}
\begin{equation}
 \begin{split}
    \mathcal{Q}_n = \left\{ \textbf{Q}^{(1)}_{n,1}(\Upsilon^{\rm H}_n),\textbf{Q}^{(1)}_{n,2}(\Upsilon^{\rm H}_n),\ldots, \textbf{Q}^{(1)}_{n,R}(\Upsilon^{\rm H}_n) \right\}.
 \label{architecture_matrix}
 \end{split}
  \vspace{-0.5em}
\end{equation}
After certain iterations, the average importance is used as the criterion and we discard the preset number of neurons with minor joint importance of its parameters.

\subsubsection{\textcolor{black}{Personalized Architecture Aggregation Based on Data}}

\begin{algorithm}[t]
\caption{Personalized architecture aggregation for generating fine headers in device cluster $\mathcal{N}_s$}
\begin{algorithmic}[1]
\STATE \textbf{Input:} $\mathcal{N}_s, \left\{\{\hat{\theta}_1^{\rm B}, \Upsilon^{\rm H}_1\},\{\hat{\theta}_2^{\rm B}, \Upsilon^{\rm H}_2\},...,\{\hat{\theta}_{|\mathcal{N}_s|}^{\rm B}, \Upsilon^{\rm H}_{|\mathcal{N}_s|}\}\right\}$, number of total iterations $T$;
\STATE Calculate $\hat{W}_s$ among devices by Eq. (\ref{wasserstain_dis}) and (\ref{normalize});
\FOR{$t=0,1,\ldots,T-1$}
    \FOR{each device $n \in \mathcal{N}_s$}
\STATE Obtain the importance set $\mathcal{Q}_n$ by Eq. (\ref{architecture_matrix});
\STATE Upload importance set $\mathcal{Q}_n$ to the edge server;
\ENDFOR
    \STATE Calculate the weighted importance sets by Eq. (\ref{customized_importance}) \\ on the edge server;
        \STATE Edge server distributes $\mathcal{Q}_n^{\prime}$ to device $n$;
    \FOR{each device $n\in \mathcal{N}_s$}
        \STATE Generate the customized header architecture by discarding unnecessary neurons based on $\mathcal{Q}_n^{\prime}$.
    \ENDFOR
\ENDFOR

\end{algorithmic}
\label{algorithm2}
\end{algorithm}

To overcome the restrictions of limited data on devices, a personalized architecture aggregation method is designed to allow the header architecture of each device to be refined by combining knowledge from other devices. As illustrated in Fig.~\ref{fig: prnue} (b), we measure the similarity of data distributions on the devices to avoid the negative impact caused by data deviation. Eventually, each device acquires a customized header that not only adapts to the local data but is also more generalized by obtaining knowledge from other devices.

Specifically, we use the Wasserstein distance to measure the differences in data distribution between different devices, which is more advantageous in capturing the geometric structure of complex distributions~\cite{li2024data}. The edge server $s_s$ calculates the similarities among devices in the assigned cluster $\mathcal{N}_s$, and generates the similarity matrix $W_s$:
\vspace{-0.5em}
\begin{equation}
    W_s = (w_{ij})_{|\mathcal{N}_s|\times |\mathcal{N}_s|},\ i,j = 1,2,...,|\mathcal{N}_s|,
 \label{wasserstain_dis}
\end{equation}
where 
$w_{ij} = \frac{1}{1 + \tilde{w}_{ij}}$ and 
$w_{ij}$ is the similarity between device $i$ and device $j$. $\tilde{w}_{ij}$ is the Wasserstein distance, i.e., 
\vspace{-0.5em}
\begin{equation*}
\tilde{w}_{ij} = \bigg( \inf_{\pi\in \Pi(\mathcal{P}(\tilde{\mathcal{D}}_i),\mathcal{P}(\tilde{\mathcal{D}}_j))}\int_{(x,y) \in X\times Y}d(x,y)^pd\pi(x,y)\bigg)^{1/p}.
\end{equation*}
$\tilde{\mathcal{D}}_i$ is a tiny portion of data randomly sampled from $\mathcal{D}_i$, and $\tilde{\mathcal{D}}_j$ is a portion of $\mathcal{D}_j$. $\mathcal{P}(\tilde{\mathcal{D}_i})$ is the distribution of features that are extracted by a pre-trained model on $\tilde{\mathcal{D}_i}$. $\Pi(\cdot, \cdot)$ denotes the set of joint distributions between two distributions. $d(x, y)$ is the distance between $x$ and $y$ calculated by L1-norm function.

To further regularize the similarity matrix $W_s$, we first turn it into a symmetric matrix and then normalize it by rows:
\vspace{-0.5em}
\begin{equation}
    \overline{W}_s=\sqrt{W_s \cdot W_s^{\mathrm{T}}}, \quad
    \hat{W}_s[i, j]=\frac{e^{\overline{W}_s[i, j]}}{\sum_{n \in \mathcal{N}_s} e^{\overline{W}_s[i, n]}}.
 \label{normalize}
\end{equation}
The final importance set for device $n$ is the
weighted convex combination of other sets with the similarity matrix $\hat{W}_s$:
\vspace{-0.5em}
\begin{equation}
   \mathcal{Q}_n^{\prime}=\sum_{i \in \mathcal{N}_s} \hat{w}_{n,i} \mathcal{Q}_i.
 \label{customized_importance}
 \vspace{-0.5em}
\end{equation}
$\mathcal{Q}_n^{\prime}$ combines knowledge from other devices, and the edge server distributes $\mathcal{Q}_n^{\prime}$ to the device $n$ for adjusting header architecture. This way, each device can have a customized header that not only matches its local data but also incorporates knowledge from the data of other devices. The detail for generating fine headers is shown in Algorithm \ref{algorithm2}.

\textcolor{black}{Our distributed \textcolor{black}{customization} approach demonstrates significant advantages over other NAS methods, such as \textit{OFA} proposed by Cai \textit{et al.}\cite{caionce}. By employing a progressive model generation and adjustment mechanism from coarse to fine granularity, we avoid uploading local data through personalized architecture aggregation of model architecture parameters, thereby preserving local data privacy. Furthermore, compared to supernet generation, our approach produces more refined models that align better with the local data distribution.}

\section{Performance Evaluation}
Since our method can serve different Transformer-based models by designing various NAS search spaces \cite{xin2021berxit, pan2024ee}, we use ViT as a representative model to evaluate the performance of \textit{ACME} for convenience in assessment.

\subsection{Experimental Setup}
\textit{\textbf{Model and dataset.}} We choose the ViT-B model \cite{liu2023survey} as the target for \textit{ACME} and use the CIFAR-100 dataset to evaluate the results. Different subsets of the dataset (with varying classes) are used as the local data for devices, achieving non-IID data distribution across devices. $s_s$ stores 10\% to 20\% of the data for $n \in \mathcal{N}_s$ as a shared dataset. 

\textit{\textbf{System settings.}} We set up 10 device clusters, each consisting of 5 virtual machines that simulate 5 individual devices. Within each cluster, the devices have similar computing and storage capabilities. Specifically, the vCPUs are configured from 3 to 7, and the model storage capacities are set to 200, 250, 300, 350, and 400 MB, respectively. Each cluster is assigned a VM to function as the edge server, while a local machine serves as the cloud server.

\textit{\textbf{Learning settings.}} We implement all methods on a Linux operating system using Python 3.9 and PyTorch 1.13.1. The entire work is trained on a Tesla V100 SXM2 32GB GPU. The LSTM used as the controller follows the configuration in \cite{zoph2018learning,10366825}, consisting of a single layer with 100 hidden units. The candidate operations in NAS include convolution operations with kernel sizes of 1, 3, and 5, identity, downsampling, average pooling, and max pooling.

\subsection{Performance Comparison with Baselines}

\begin{table}[t]
    \centering
    \caption{Analysis of cost-efficiency for different architectures}
    \vspace{-1em}
     \begin{center}
    \begin{tabular}{|c|c:c|c:c|}
        \hline
        \multirow{2}{*}[-0.5ex]{$N$} & \multicolumn{2}{c|}{{Search Space ($10^3$)}} & \multicolumn{2}{c|}{{Upload Data (MB)}} \\
        \cline{2-5}
         & {CSs} & {Ours}& {CSs} & {Ours} \\
        \hline
        10 & 1695 & \textbf{17.2} & 1610 & \textbf{96.6} \\
        20 & 3300 & \textbf{33.4} & 3220 & \textbf{193.2}  \\
        30 & 4050 & \textbf{41.1} & 4830 & \textbf{289.8}  \\
        40 & 6600 & \textbf{66.8} & 6440 & \textbf{386.4}  \\
        \hline
    \end{tabular}
    \end{center}
    \vspace{-1em}
    \label{table}
\end{table}

\begin{figure}[t]
	\centering

   \setlength{\belowcaptionskip}{-0.3em} 
	\includegraphics[width=\linewidth]{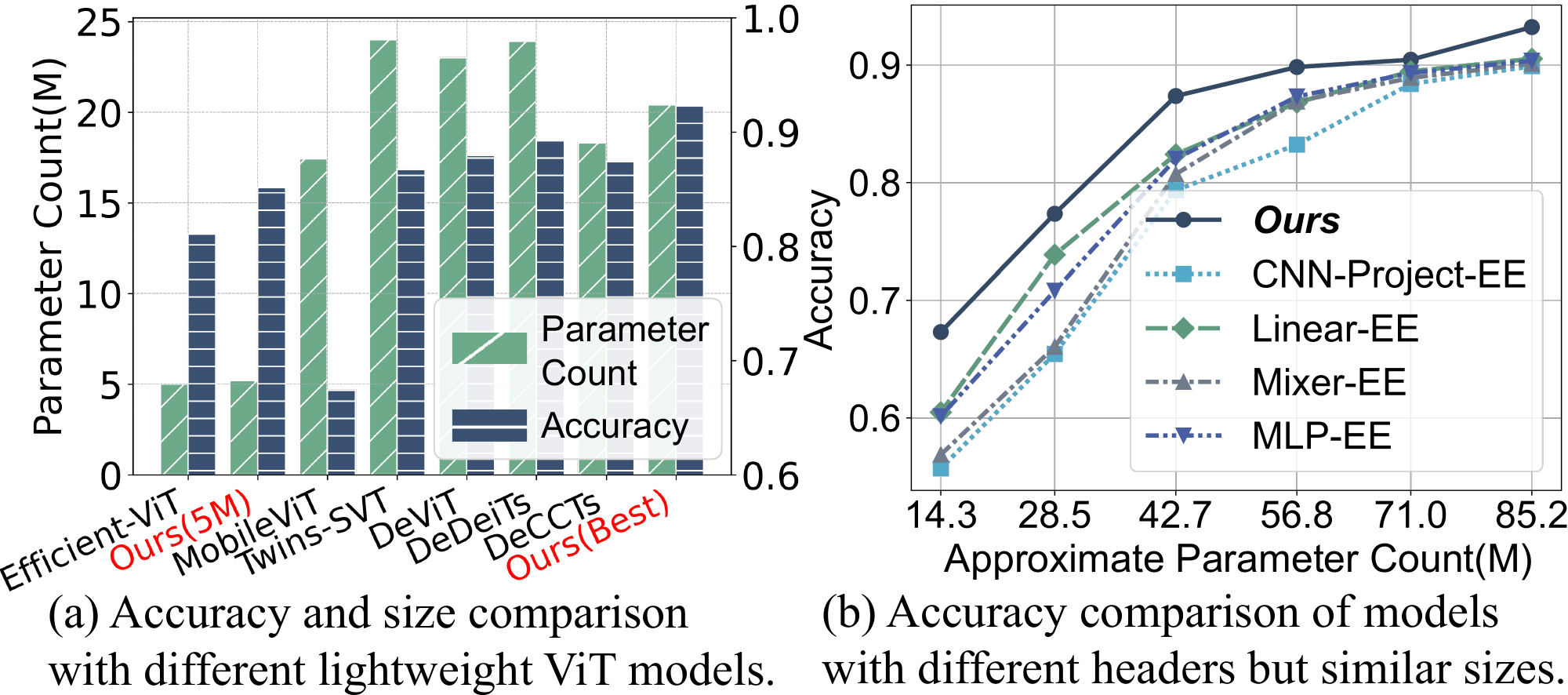}
	\caption{Performance comparison with baselines on CIFAR-100.}
	\label{fig: baselines}
  \vspace{-1em}
\end{figure}

\begin{figure}[t]
	\centering

	\includegraphics[width=\linewidth]{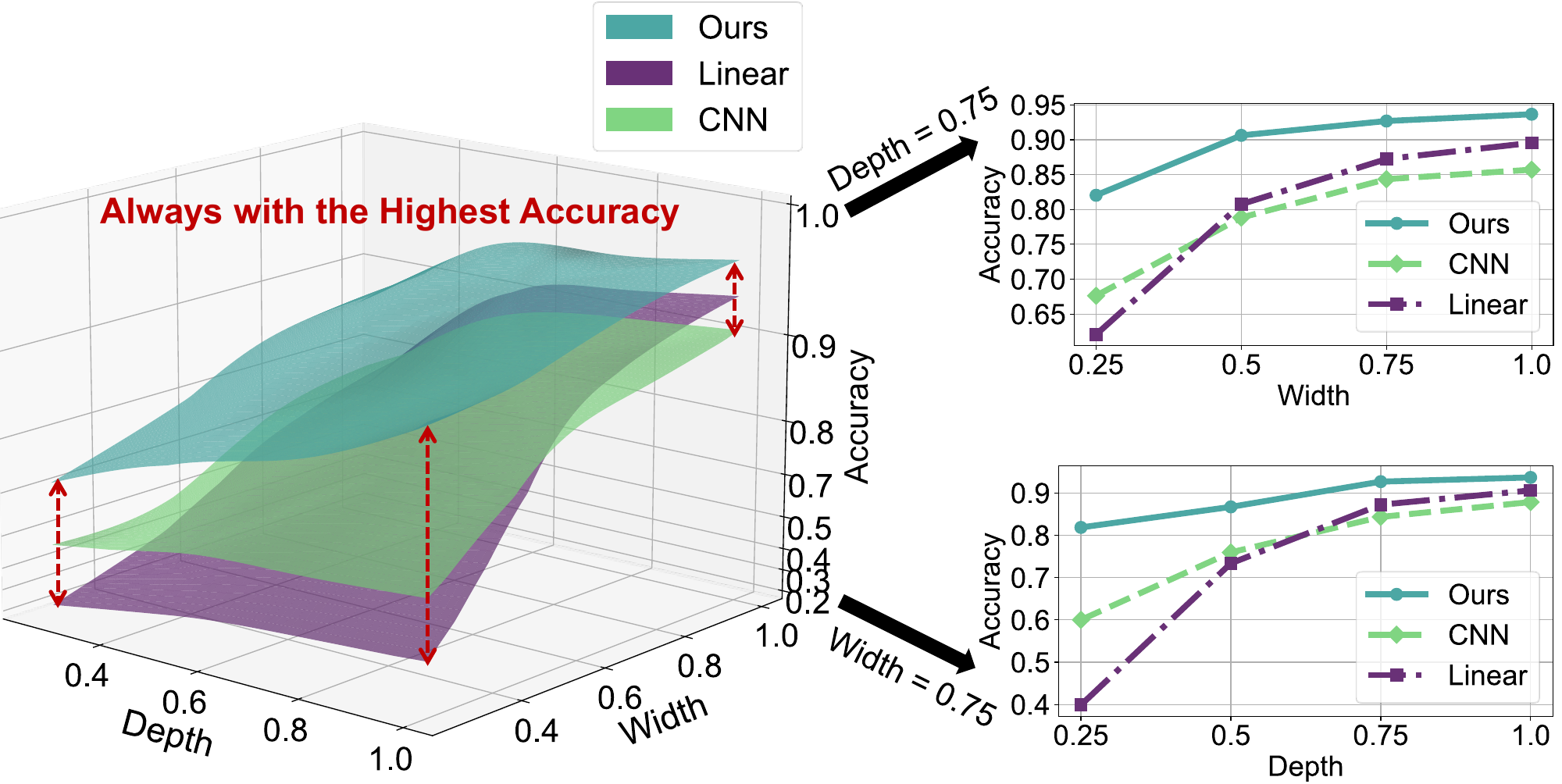}
	\caption{
 Accuracy of different headers applied to varying backbones.
 }
	\label{fig: 3dgraph}
  \vspace{-1em}
\end{figure}

\subsubsection{Analysis of System Cost-Efficiency}\label{experi1}
\textcolor{black}{The results shown in Table \ref{table} highlight the superior cost-efficiency of our proposed distributed system compared to traditional CSs. \textit{ACME} employs a progressive customization approach, moving from the backbone to the header, where NAS is utilized only during the coarse-grained header generation. This strategy reduces the search space to approximately 1\% of that in CSs, enabling edge servers to execute the algorithm. Moreover, model customization on the cloud and edge servers does not require local datasets, resulting in a significantly lower data upload volume compared to CSs that rely on local datasets for training—on average, decreasing to 6\% of CSs. These findings demonstrate that using a distributed system for model customization offers high cost-efficiency.}

\subsubsection{Learning Performance}
We use the average accuracy of all devices to demonstrate the performance of the model. As shown in Fig. \ref{fig: baselines} (a), we compare \textit{ACME} applied to ViT-B with other lightweight ViT model baselines:
($i$) \textit{\textbf{Efficient-ViT}} \cite{xie2023efficient}, ($ii$) \textit{\textbf{MobileViT}} \cite{mehtamobilevit}, ($iii$) \textbf{\textit{Twins-SVT}}\cite{chu2021twins},  ($iv$) \textbf{\textit{DeViT}, \textit{DeDeiTs}, and \textit{DeCCTs}} \cite{xu2023devit}. We set the maximum device storage constraint to 25M. Our method's best model performs well on both indicators, with nearly 10\% accuracy improvement. 
Compared to ($iv$), which also customizes the backbone and header separately, our method achieves nearly 5\% accuracy improvement with only 85.3\% of their parameters. Compared to ($iii$), we achieve 15\% fewer parameters and 5.62\% higher accuracy. Even compared to a tiny model like ($i$), our method shows an accuracy improvement of 4.07\% for similar model sizes. For heterogeneous attributes of devices, our method can achieve the optimal trade-off models on devices with different storage constraints by adjusting the lightweight degree of the backbone and automatically designing a header.
 
As shown in Fig. \ref{fig: baselines} (b), we illustrate the accuracy differences between the four headers designed in \cite{bakhtiarnia2021multi} and the headers generated by our method when applied to backbones of similar model sizes. To demonstrate that our method can design headers through NAS to achieve higher model accuracy compared to traditional headers with the same backbone, we fix the backbone width to 1. The result shows that our method improves the model accuracy by an average of 9.02\% on small backbones and nearly 3\% on large backbones. Compared to traditional headers, using NAS for automatic design can more effectively obtain model architectures that are better suited to the backbone, thereby enhancing model performance.

To further explore the superiority of NAS-generated headers, we compare different backbone architectures, as shown in Fig. \ref{fig: 3dgraph}. From the left figure, it is evident that our method significantly outperforms traditional headers across various backbone architectures. For a more detailed analysis, we examine points where the depth or width is 0.75. It can be observed that when the backbone model is simple, a more complex header can better compensate for the backbone's limited feature extraction capability, making CNN-based headers more effective than Linear ones. Conversely, when the backbone architecture is complex, a simpler header architecture can more effectively leverage the backbone's capabilities. Using our method, we can automatically adjust the complexity of the header, ensuring optimal performance across different backbone architectures.
\begin{figure}[t]
	\centering
   \setlength{\abovecaptionskip}{-0.1em} 
   \setlength{\belowcaptionskip}{-0.1em} 
	\includegraphics[width=0.97\linewidth]{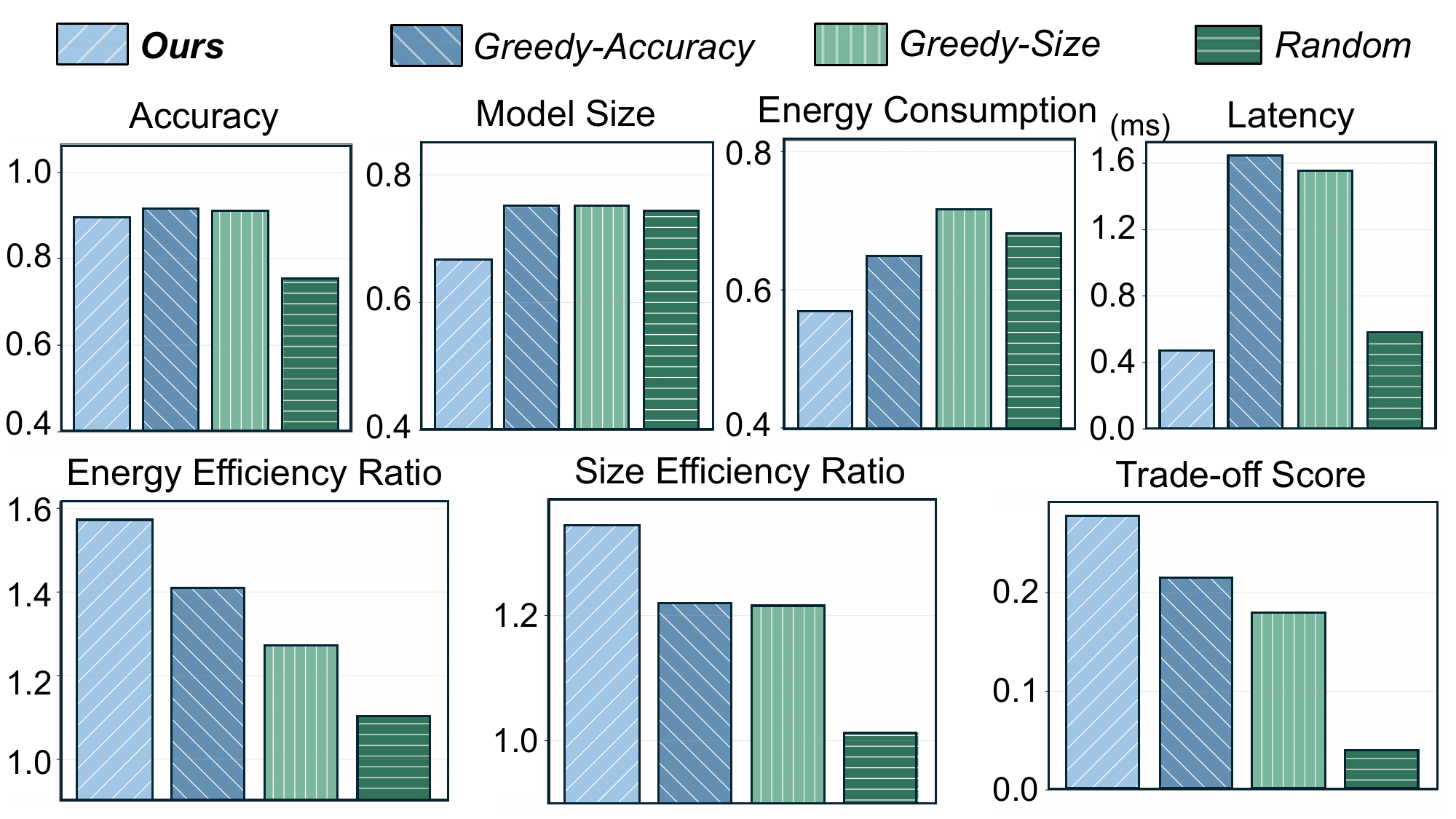}
	\caption{Comparison of models under different matching methods.}
	\label{fig: matching}
  \vspace{-1.5em}
\end{figure}

\subsubsection{Analysis of Model and Device Attributes Matching}\label{experi3}
As shown in Fig. \ref{fig: matching}, we compare our model matching method with three common methods for matching lightweight models: \textbf{Greedy-Accuracy} (selecting the highest accuracy model that can be deployed)\cite{howard2017mobilenets}, \textbf{Greedy-Size} (selecting the largest model that can be deployed)\cite{gordon2018morphnet}, and \textbf{Random} matching. We evaluate these methods from multiple perspectives. We use the \textbf{Energy Efficiency Ratio} to measure the accuracy that can be achieved per unit of energy consumption, which is the ratio of model accuracy to energy consumption. Similarly, we also define the \textbf{Size Efficiency Ratio} to evaluate the accuracy achieved per unit of model size. Lastly, we define the \textbf{Trade-off Score} as $\mathcal{L}_n(\theta_n, \mathcal{D}_n)+E_n(\theta_n)+\zeta(\theta_n)$ to assess the overall cost-efficiency of the models\cite{kim2006adaptive}. It is evident that our method, after constructing the Pareto Front, has a selection latency comparable to the Random method, reducing selection latency by 71.2\% compared to the Greedy methods. Additionally, our method achieves the highest energy and model size efficiency and improves the final Trade-off Score by at least 28.9\%. This demonstrates that our matching method can find the optimal model match in a short time.

\subsubsection{Analysis of Model and Device Data Matching}\label{experi4}
\begin{figure}[t]
	\centering
        \setlength{\abovecaptionskip}{-0.1em} 
        \setlength{\belowcaptionskip}{-0.1em} 
	\includegraphics[width=\linewidth]{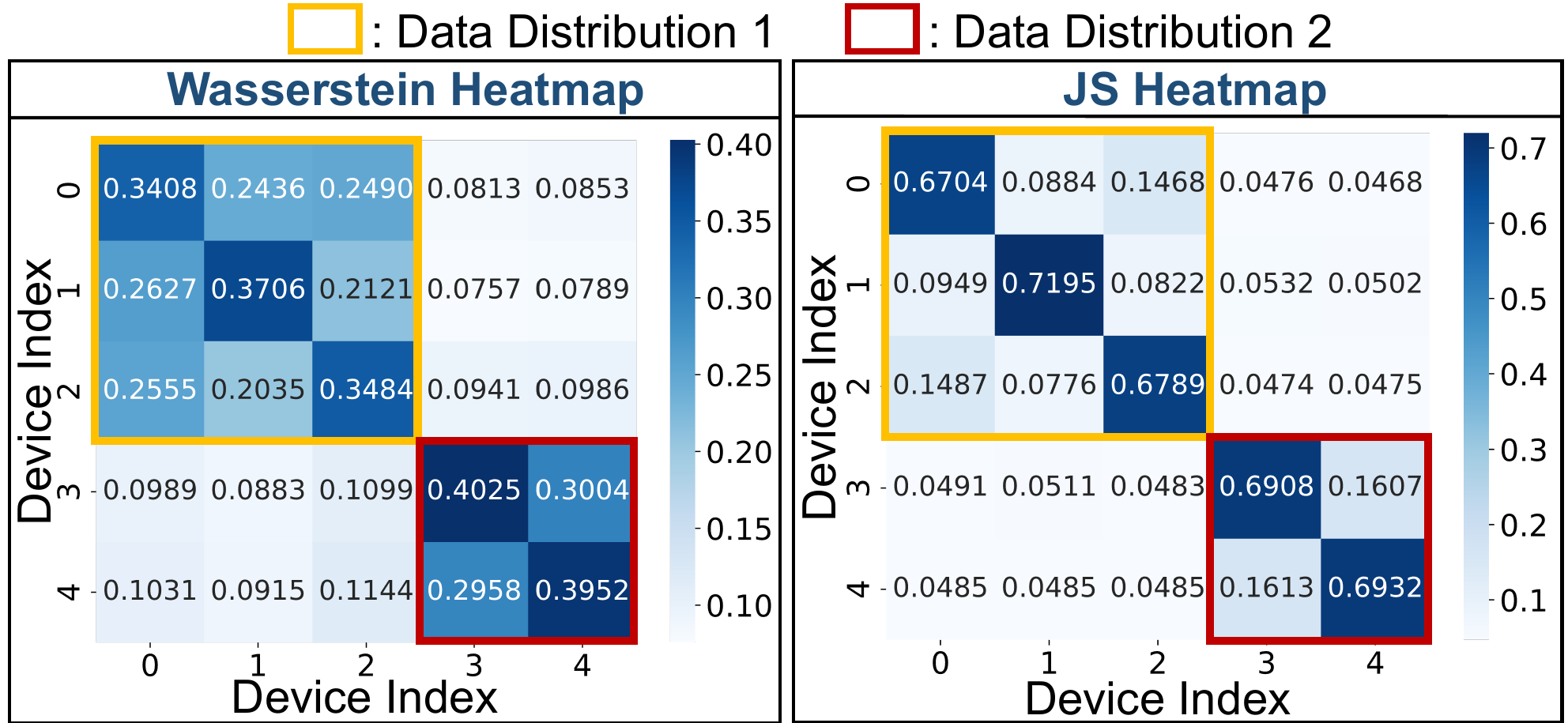}
        \vspace{-1em}
	\caption{Comparison of the feature extraction ability of different distances.}
	\label{fig: sim_heat}
\end{figure}

In order to better match device data, our method aims to select devices with similar data distributions for collaboration. We employ both Wasserstein distance and Jensen-Shannon (JS) divergence to generate similarity matrices for devices with specific data distributions (where devices 0 to 2 share the same distribution, and devices 3 and 4 share a different distribution) for visual analysis, as shown in Fig.~\ref{fig: sim_heat}. Compared to the JS divergence, the Wasserstein distance in \emph{ACME} more accurately captures the complex data relationships between devices.

\begin{figure}[t]
	\centering
        \includegraphics[width=\linewidth]{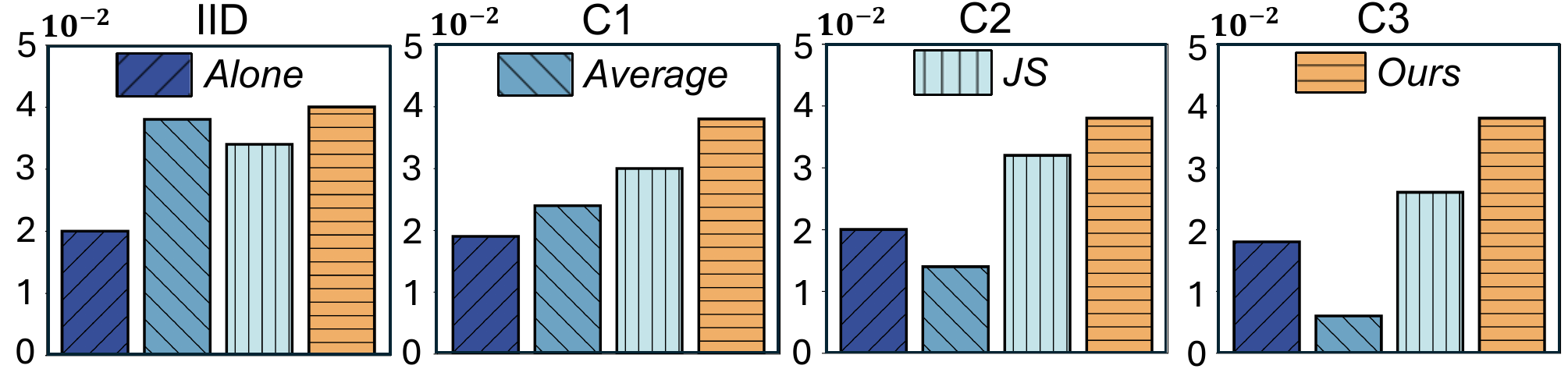}
        \vspace{-2em}
	\caption{The accuracy improvement under four data distributions.}
	\label{fig: emd_acc}
  \vspace{-1em}
\end{figure}

To demonstrate the superiority of personalized architecture aggregation by \textit{ACME}, we compare the accuracy improvement of four methods on both IID and non-IID data distributions (including C1, C2, and C3 with increased confusion levels) across a uniform cluster of devices. \textbf{Alone} refines the model with a local importance set. \textbf{Avg} refines the model by averaging the importance sets among devices. \textbf{JS} uses JS divergence to measure the similarity among devices. We use the improved accuracy of the original model on the device as the evaluation metric. As shown in Fig.~\ref{fig: emd_acc}, \textit{ACME} exhibits the highest accuracy across all data distributions. Compared to the original model, four methods by adjusting model architecture based on the device's local data result in varying degrees of accuracy improvement. Meanwhile, inter-device collaboration can further improve accuracy under IID data distribution. However, as data complexity increases, the \textbf{Avg} method, which ignores inter-device correlations, loses its advantages. \emph{ACME}'s accuracy improvement is more significant due to its precise capture of device similarity using Wasserstein distance.

\subsection{Impact of Key Parameters on \textit{ACME}}
\begin{figure}[t]
	\centering
    \setlength{\abovecaptionskip}{-0.2em} 
   \setlength{\belowcaptionskip}{-0.2em} 
	\includegraphics[width=\linewidth]{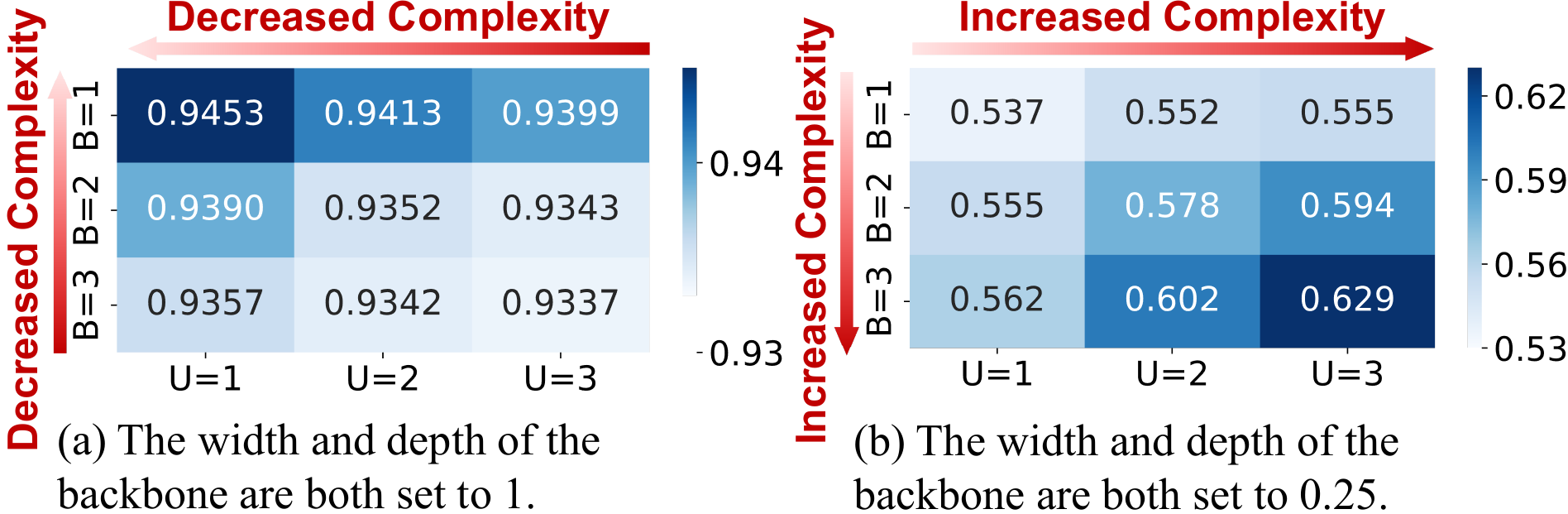}
	\caption{Impact of block and layer count in header architecture.}
	\label{fig: nas_hyper}
  \vspace{-0.5em}
\end{figure}

\textbf{\textit{Complexity of the header search space. }}As shown in Fig. \ref{fig: nas_hyper} (a), when the backbone model is large, its architecture is already sufficient to extract the features of the input image. Therefore, a relatively simple header architecture is adequate for achieving high accuracy for downstream tasks. Conversely, if the header is too complex, it may lead to feature loss and reduce the model's accuracy. When the backbone architecture is straightforward, the model needs a more complex header to provide feature extraction capabilities. Therefore, Fig. \ref{fig: nas_hyper} (b) illustrates a phenomenon where model accuracy improves as $B$ and $U$ increase.

\subsection{Auxiliary Validation}
\begin{figure}[t]
	\centering
   \setlength{\abovecaptionskip}{-0.2em} 
   \setlength{\belowcaptionskip}{-0.2em} 
	\includegraphics[width=\linewidth]{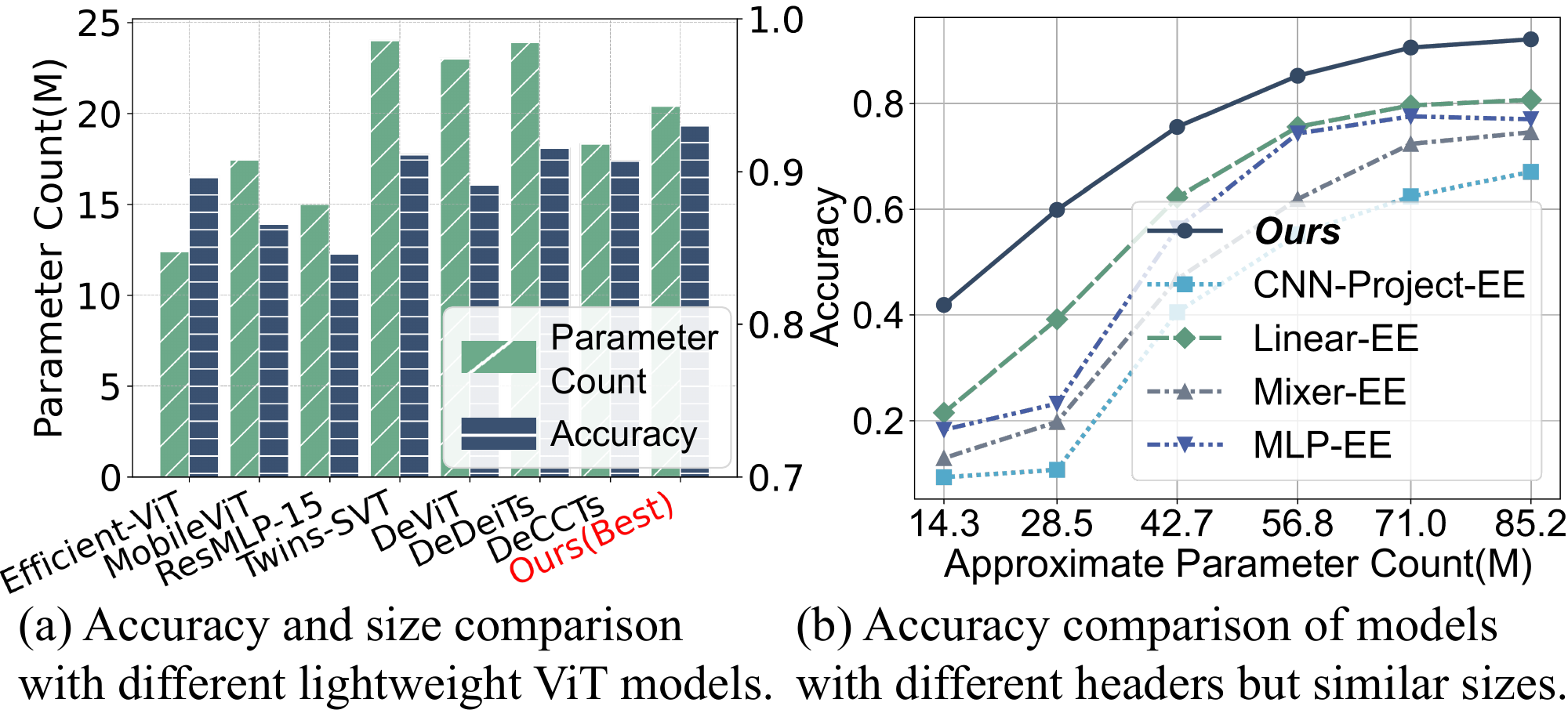}
	\caption{Performance comparison with baselines 
 on Stanford Cars.}
	\label{fig: stan_baseline}
  \vspace{-1em}
\end{figure}

To further validate \textit{ACME}, we conduct tests on the Stanford Cars dataset\cite{krause20133d}. As shown in Fig. \ref{fig: stan_baseline} (a), \textit{ACME} also achieves the performance-optimal model under the 25M storage constraint on this dataset. The model we obtained shows an average accuracy improvement of 3.94\% under storage constraints. Fig. \ref{fig: stan_baseline} (b) shows that when modifying the header, our method achieves significantly better model accuracy on more complex datasets, with an average accuracy improvement of 14.43\% across different model sizes. \textit{ACME}'s approach of automatically generating model architectures based on the dataset achieves a balance between model lightweight and accuracy. The results also demonstrate the adaptability of this method across different datasets.
  \vspace{-0.5em}

\section{Related Work}

\textit{\textbf{Deployment of Transformer-based models on devices.}} Many recent studies aim to use lightweight Transformer-based models to meet the requirements of device deployment. \textit{DynaBERT} and \textit{DeViT} achieve lightweight versions of BERT and ViT for device deployment by realizing dynamic width and depth, respectively\cite{hou2020dynabert, xu2023devit}. \textit{MobileViT} combines CNN with ViT to provide a lightweight model suitable for mobile devices\cite{mehtamobilevit}. Building on this, Xie \textit{et al.} propose \textit{Efficient-ViT}, which processes local information with CNN and global information with ViT\cite{xie2023efficient}. Chu \textit{et al.} introduce conditional positional encodings and spatially separable self-attention to reduce ViT's complexity\cite{chu2021twins}. While these methods compress models for deployment on devices, they overlook the need for refining models to suit heterogeneous devices. Some NAS methods have achieved model customization, such as \textit{FedNAS} and \textit{CFDNAS}\cite{yuan2022resource,zhang2022toward}, which support model adaptation for heterogeneous devices. However, these approaches are not suitable for Transformer-based models, as the large model size leads to prohibitively high computational complexity for NAS.

\textit{\textbf{Multi-exit and early-exit through header design.}} In this field, studies often improve model inference speed through multi-exit or early-exit methods. Bakhtiarnia \textit{et al.} design seven different header architectures for ViT to enable dynamic inference by leveraging the multi-exit architecture of deep neural networks\cite{bakhtiarnia2021multi}. Xu \textit{et al.} propose the \textit{LGViT} framework, which includes two different header architectures to balance inference efficiency and accuracy\cite{xu2023lgvit}. This early-exit inference technique can significantly reduce the model size required for inference, thereby achieving a lightweight model. However, most works do not focus on the potential of this technique for large models to enable their deployment on devices.

\section{Conclusion}

\textcolor{black}{In summary, this paper has introduced \textit{ACME}, an approach for achieving adaptive model customization through a bidirectional single-loop distributed system. The core design of this approach involves hierarchical processing of the backbone and header, leveraging distributed systems to progressively customize models from coarse to fine granularity. This has addressed the challenge of mismatches between models and devices with heterogeneous attributes and data. Experiments conducted on two datasets have demonstrated that \textit{ACME} effectively customizes cost-efficient models for devices under various performance and storage constraints. Compared to baselines with similar model sizes, \textit{ACME} has achieved higher accuracy by automatically generating header architectures and fine-tuning them based on local data.
}

\section{Acknowledgement}
This research is supported by Technologies for Efficient Crowd Intelligence Understanding and Situation Deduction in Intelligent Connected Vehicle Environments, under Grant No. F2024201070, Tianjin Natural Science Foundation General Project No. 23JCYBJC00780, National Natural Science Foundation of China under Grant No. U23B2049, the Tianjin Xinchuang Haihe Lab under Grant No. 22HHXCJC00002.

\bibliographystyle{IEEEtran}
\bibliography{Ref}

\end{document}